\documentclass[12pt,oneside,british,french]{book}
\usepackage[T1]{fontenc}
\usepackage[latin1]{inputenc}
\usepackage[letterpaper]{geometry}
\usepackage{array}
\usepackage{verbatim}
\usepackage{graphicx}
\usepackage{setspace}
\makeatletter

\newcommand{\noun}[1]{\textsc{#1}}
\providecommand{\tabularnewline}{\\}

\usepackage{udsbib}

\makeatother

\usepackage{babel}
\makeatletter
\addto\extrasfrench{%
   \providecommand{\og}{\leavevmode\flqq~}%
   \providecommand{\fg}{\ifdim\lastskip>\z@\unskip\fi~\frqq}%
}

\makeatother
\begin{document}
\selectlanguage{british}%
\begin{titlepage}\centering{\includegraphics[width=0.4\columnwidth]{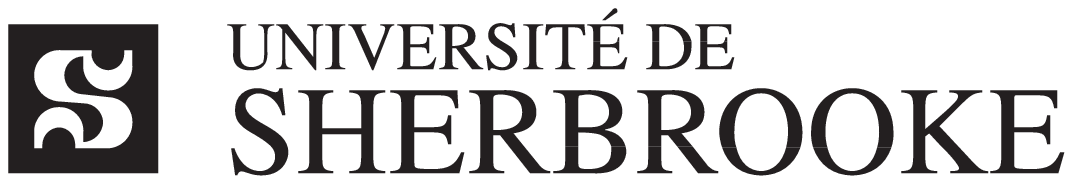}

\selectlanguage{french}%
Faculté de génie

Génie électrique et informatique

\vfill{}

{\Large{}Extension spectrale d'un signal de parole de la bande téléphonique
à la bande AM}{\Large \par}

\vfill{}

Mémoire de maîtrise

Specialité: génie électrique

\vfill{}

Jean-Marc Valin

\vfill{}

}

Sherbrooke (Québec) Canada\hfill{}décembre 2001

\selectlanguage{british}%
\end{titlepage}\thispagestyle{empty}\cleardoublepage

\selectlanguage{french}%
\pagenumbering{roman}

\chapter*{Résumé}

Le présent mémoire propose un système d'extension de la bande permettant
de produire un signal en bande AM à partir d'un signal de parole en
bande téléphonique. L'objectif est donc de reconstruire le signal
en bande AM avec une qualité sonore se rapprochant le plus possible
de la référence en bande AM.

L'extension est effectuée de façon indépendante pour les hautes fréquences
et les basses fréquences. L'extension des hautes fréquences utilise
le modèle filtre-excitation, ce qui divise le problème en deux parties:
l'extension de l'excitation et de l'enveloppe spectrale. L'extension
de l'excitation est réalisée dans le domaine temporel par une fonction
non linéaire, alors que l'extension de l'enveloppe spectrale s'effectue
dans le domaine cepstral par un perceptron multi-couches. L'extension
de la bande basse utilise le modèle sinusoïdal. L'amplitude des sinusoïdes
est aussi estimée par un perceptron multi-couches. 

Les résultats obtenus montrent que la qualité sonore après extension
est supérieure à celle de la bande téléphonique. Toutefois, on note
une importante différence de perception entre différents auditeurs.

Certaines techniques développées pour le projet d'extension de la
bande présentent un certain intérêt pour le domaine du codage de la
parole. L'extension de l'excitation est l'une d'entre elles et fait
l'objet d'une étude plus approfondie.

\chapter*{Remerciements}

Je voudrais d'abord remercier mon directeur de recherche et professeur,
Roch Lefebvre pour son aide précieuse tout au long de ce projet. Je
voudrais ensuite remercier la compagnie VoiceAge, ainsi que le Conseil
de recherches en sciences naturelles et en génie (CRSNG) pour leur
support financier au cours de ma maîtrise.

Je suis reconnaissant à Dominic Létourneau pour m'avoir aidé à plusieurs
reprises. Je remercie ma fiancée Nathalie, pour sa patience et son
support malgré les nombreuses journées de travail s'achevant souvent
à 4h du matin! Je veux remercier mes parents, Laila et Michel qui
m'ont supporté tout au long de mes études.

J'aimerais enfin remercier les nombreux volontaires qui ont accepté
de participer aux évaluations subjectives. 

\clearpage

\tableofcontents{}

\clearpage

\listoffigures

\clearpage

\listoftables

\clearpage

\chapter*{Lexique}

\vspace{0.375cm}

\begin{center}
\begin{tabular}{p{1.5cm}p{20cm}}
\textbf{\noun{CELP}} & Code-Excited Linear Prediction\tabularnewline
\textbf{CODEC} & COdeur DÉCodeur\tabularnewline
\textbf{DCT} & Discrete Cosine Transform (transformée en cosinus discrète)\tabularnewline
\textbf{\noun{DSP}} & Digital Signal Processor\tabularnewline
\textbf{FFT} & Fast Fourier Transform (transformée de Fourier rapide)\tabularnewline
\textbf{\noun{LPC}} & Linear Prediction Coefficients (coefficients de prédiction linéaire)\tabularnewline
\textbf{MOS} & Mean Opinion Score\tabularnewline
\textbf{MMSE} & Minimum Mean Square Error (erreur quadratique minimale)\tabularnewline
\textbf{pitch} & Fréquence fondamentale de la voix\tabularnewline
\textbf{QV} & Quantification vectorielle\tabularnewline
\textbf{RELP} & Residue-Excited Linear Prediction\tabularnewline
\textbf{RSB} & Rapport signal sur bruit (aussi SNR: Signal to Noise Ratio)\tabularnewline
\end{tabular}
\par\end{center}

\vspace{0.375cm}

\chapter{Introduction}

\pagenumbering{arabic}

\section{Description du problème}

La plupart des applications de transmission vocale actuelles transmettent
la parole dans la \og bande téléphonique \fg{}, soit de $200\,\mathrm{Hz}$
à $3500\,\mathrm{Hz}$. Toutefois, afin d'améliorer la qualité de
la voix, on a de plus en plus recours à des systèmes transmettant
la voix dans la bande AM, soit de $50\,\mathrm{Hz}$ à $7000\,\mathrm{Hz}$.
Un signal dans la bande AM est souvent appelé signal \og large bande \fg{}. 

Alors qu'une fréquence d'échantillonnage de $8\,\mathrm{kHz}$ est
suffisante pour transmettre un signal en bande téléphonique, la bande
AM nécessite une fréquence d'échantillonnage de $16\,\mathrm{kHz}$.
Malheureusement, le réseau téléphonique actuel a été conçu pour fonctionner
à une fréquence d'échantillonnage de $8\,\mathrm{kHz}$ et il est
presque impensable de modifier tout le matériel du réseau téléphonique
pour que ce dernier fonctionne à $16\,\mathrm{kHz}$.

C'est pour cette raison qu'il existe un besoin pour un système qui
permettrait de produire un signal large bande à partir du signal correspondant
dans la bande téléphonique. Ce système ferait en sorte de modifier
uniquement les récepteurs téléphoniques, sans modifier le réseau lui-même.
Toutefois, pour des applications de téléphonie numérique, il peut
être préférable de \og dépenser \fg{} quelques bits supplémentaires
afin d'obtenir une qualité de son se rapprochant encore plus de la
qualité \og large bande \fg{}. On pourra alors coder la partie de
l'information qui n'a pu être restaurée en ajoutant un faible débit
à un codeur de parole en bande téléphonique.

\section{Bande téléphonique et bande AM}

La majeure partie de l'information contenue dans un signal de parole
est comprise entre $50\,\mathrm{Hz}$ et $7\,\mathrm{kHz}$, soit
dans la bande AM. Le fait de transmettre uniquement la bande téléphonique
($200\:\mathrm{Hz}$ à $3,5\:\mathrm{kHz}$) du signal affecte considérablement
la qualité perçue par l'utilisateur tout en diminuant légèrement l'intelligibilité.
Chacune des bandes de fréquence perdue lors du filtrage dans la bande
téléphonique affecte d'une manière différente le signal de parole
tel que perçu par l'utilisateur. 

La partie basses fréquences perdue est surtout liée à l'impression
de \og présence \fg{} lors de la communication, surtout lorsque
le locuteur est un homme. En effet, les fréquences entre $50\,\mathrm{Hz}$
et $200\,\mathrm{Hz}$ sont surtout perçues comme des vibrations qui
indiquent normalement que le locuteur est proche.

La partie hautes fréquences, quant à elle, donne une impression de
\og clarté \fg{} au signal de parole. Un signal de parole en bande
téléphonique perd beaucoup de cette clarté, surtout pour des locuteurs
féminins. De plus, c'est surtout par les hautes fréquences que l'oreille
arrive à discriminer les fricatives (/s/, /f/, /$\int$/) entre elles.
Pour cette raison, la partie hautes fréquences contribue à l'intelligibilité
de la parole.

\section{Solution proposée}

Contrairement à d'autres travaux qui ne s'intéressent qu'à la bande
basse ou la bande haute, ce projet consiste à faire l'extension des
deux bandes, soit de $50\,\mathrm{Hz}$ à $200\,\mathrm{Hz}$ et de
$3500\,\mathrm{Hz}$ à $8000\,\mathrm{Hz}$. De plus, le signal de
parole utilisé en entrée du système est filtré de la même manière
qu'un signal ayant passé par le réseau téléphonique.

En plus de l'objectif premier du projet qui est de faire un système
d'extension complet, il est aussi question de l'application de certaines
techniques développées pour ce projet au domaine plus général du codage
large bande. En effet, on sait qu'une technique permettant de prédire
une certaine variable permet de diminuer la quantité d'information
nécessaire à la quantification de cette même variable (pourquoi coder
l'information qui peut être obtenue autrement?).

\section{Caractéristiques recherchées\label{sec_objectifs_initiaux}}

Afin d'être utile dans une application de téléphonie, il est important
que le système d'extension de la bande conçu possède certaines caractéristiques.
Celles-ci sont: la qualité du son, une complexité raisonnable et un
faible délai algorithmique.

\subsection{Qualité sonore}

La qualité sonore est, bien entendu, la première caractéristique recherchée,
puisque c'est justement pour augmenter la qualité de la parole que
l'on désire passer de la bande téléphonique à la bande AM. Le signal
de parole traité doit être plus agréable à écouter que le signal original
en bande téléphonique et sa qualité doit se rapprocher le plus possible
de celle du signal \og original \fg{} large bande.

\subsection{Complexité}

La complexité est une caractéristique très importante de tout système
de traitement de la voix. En effet, ces systèmes sont, la plupart
du temps, implantés sur des DSP ayant une capacité de traitement et
une mémoire limitées. Bien qu'aucune mise en oeuvre sur DSP ne soit
proposée, il faudra tout de même tenir compte de contraintes de complexité
\og raisonnables \fg{}.

\subsection{Délai algorithmique}

Afin de pouvoir être utilisé lors d'une conversation en temps réel,
un algorithme de traitement de la parole doit avoir un délai qui ne
soit pas perceptible par l'utilisateur. Ceci exclut certains traitement
non causals qui nécessitent de connaître une partie importante du
signal à venir. Pour des applications de voix en temps réel, on accepte
généralement un délai de traitement ne dépassant pas $100\,ms$.

\section{Organisation du mémoire}

Comme le présent projet requiert une certaine connaissance du signal
de parole et des outils d'analyse de ce signal, cette partie sera
traitée au chapitre \ref{sec_signal_et_modelisation}. Suivra ensuite
au chapitre \ref{sec_overview} un aperçu global du système proposé.
Les chapitres \ref{sec_ext_HF} et \ref{sec_ext_BF} traiteront respectivement
des algorithmes utilisés pour reconstruire les hautes et les basses
fréquences. Le chapitre \ref{sec_resultats} sera consacré aux résultats
obtenus, suivi par une discussion au chapitre \ref{sec_conclusion}.

\chapter{Signal de parole et modélisation\label{sec_signal_et_modelisation}}

\section{Caractéristiques du signal de parole\label{sec_caract_parole}}

\begin{figure}[!t]
\begin{centering}
\includegraphics[width=1\columnwidth]{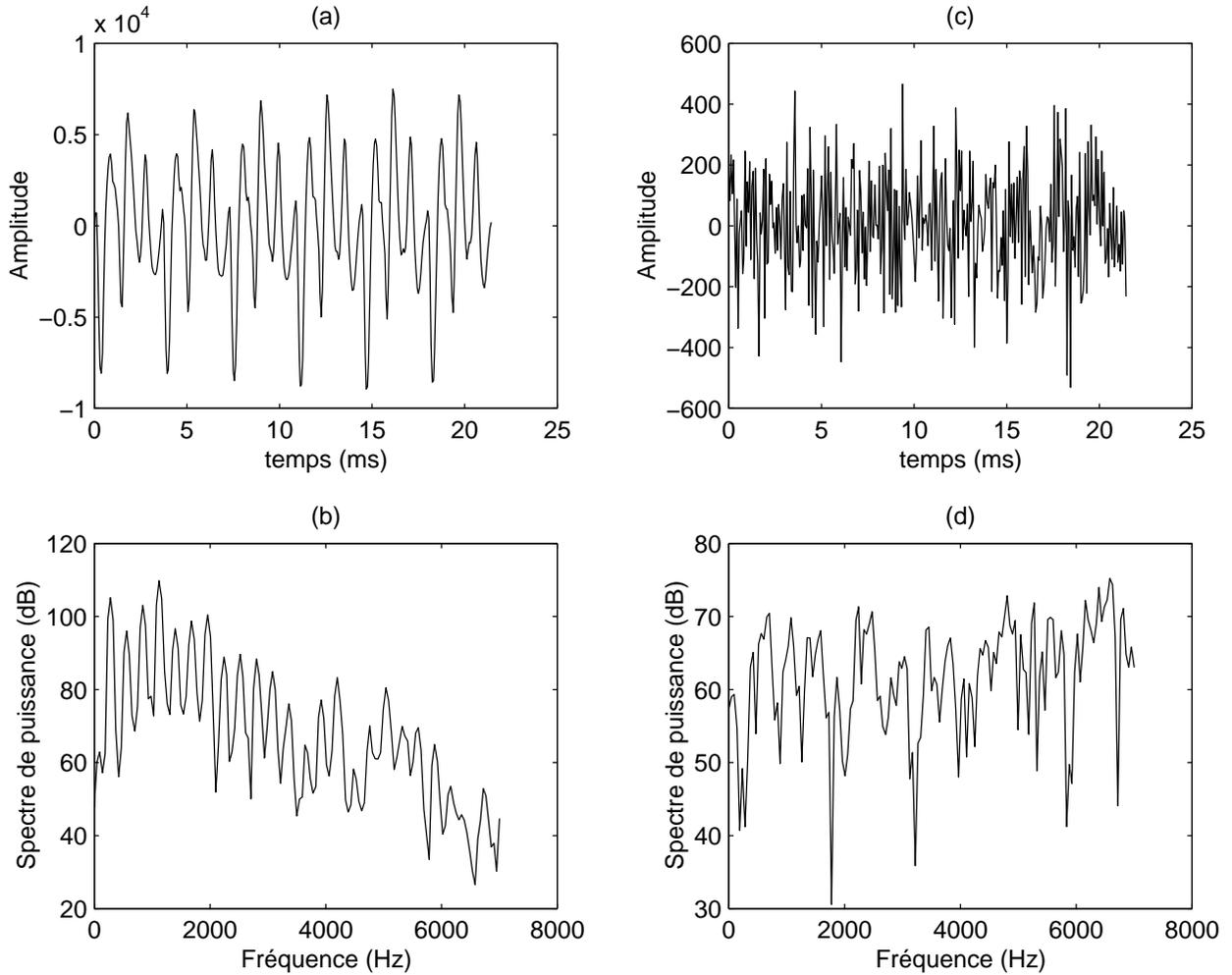}
\par\end{centering}

\caption[Signaux voisés et non voisés]{Illustration des différences entre un phonème voisé et un phonème non voisé. (a) Le phonème voisé /a/ (voyelle) dans le domaine temporel et (b) fréquentiel. (c) Le phonème non voisé /s/ (fricative) dans le domaine temporel et (d) fréquentiel.}\label{fig_voises_non_voises}
\end{figure}

La parole est produite lorsque l'air, poussé hors des poumons, passe
par les cordes vocales et le conduit vocal pour produire un son. Les
modes de production des sons diffèrent grandement, ce qui permet une
grande variété de phonèmes. On peut diviser ces phonèmes en deux classes:
voisés et non voisés. Les voyelles sont des exemples de sons voisés,
alors que les fricatives sont des exemples de sons non voisés. La
figure \ref{fig_voises_non_voises} montre, dans le domaine temporel
et fréquentiel, une voyelle (/a/) et une fricative (/s/). On remarque
que le signal correspondant à une voyelle est périodique, ce qui s'explique
par le fait que la glotte s'ouvre et se referme à intervalles réguliers.
Pour les fricatives, la glotte reste grande ouverte, ce qui explique
l'absence de périodicité. Les phonèmes voisés se distinguent dans
le domaine spectral par une structure fine harmonique, alors que la
structure fine des phonèmes non voisés est stochastique.

\begin{figure}[!t]
\begin{centering}
\includegraphics[width=1\columnwidth]{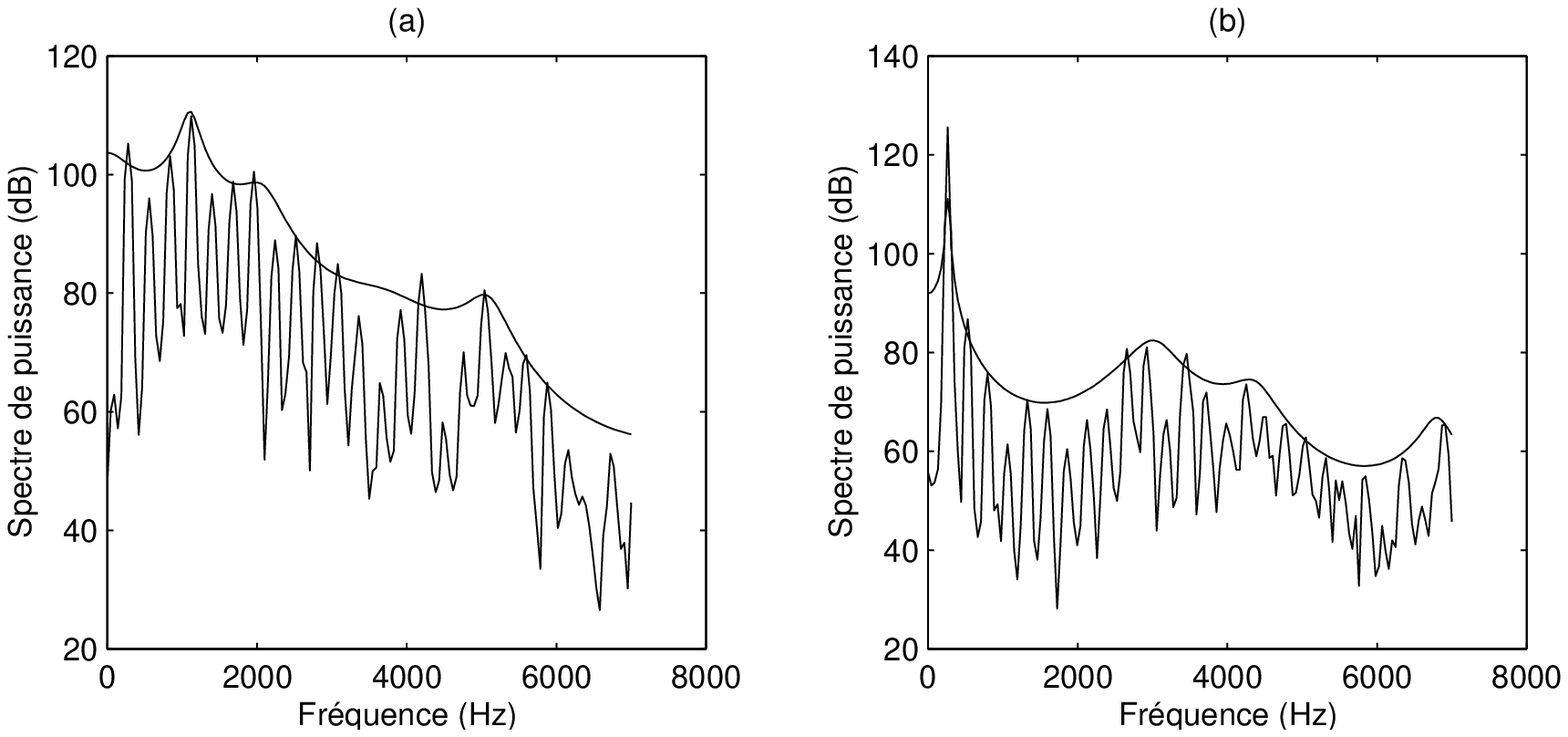}
\par\end{centering}

\caption[Discrimination des voyelles par les formants]{Illustration des différences entre les formants de différentes voyelles. (a) Le phonème /a/. (b) Le phonème /i/.}\label{fig_formants}
\end{figure}

En plus d'être différents par leur voisement, les phonèmes se distinguent
par la forme générale de leur spectre, soit l'enveloppe spectrale.
Les résonances présentes dans l'enveloppe sont appelées \og formants \fg{}
et permettent de discriminer les voyelles entre elles. On peut voir
à la figure \ref{fig_formants} la différence entre les formants d'un
/a/ et ceux d'un /i/.

Le signal de parole peut donc être modélisé simplement comme une source
d'excitation (sortie des cordes vocales) filtrée par un filtre résonant
(conduit vocal) représentant l'enveloppe spectrale. On connaît cette
représentation de la parole comme modèle \og filtre-excitation \fg{}.
Afin de modéliser la réponse en fréquence du conduit vocal, on utilise
de façon générale un filtre tout-pôles, dont les coefficients évoluent
dans le temps. Si le filtre tout-pôles représente bien l'enveloppe
spectrale du signal, c'est dire que le spectre de l'excitation doit
être généralement plat. 

\begin{figure}[!t]
\begin{centering}
\includegraphics[width=1\columnwidth]{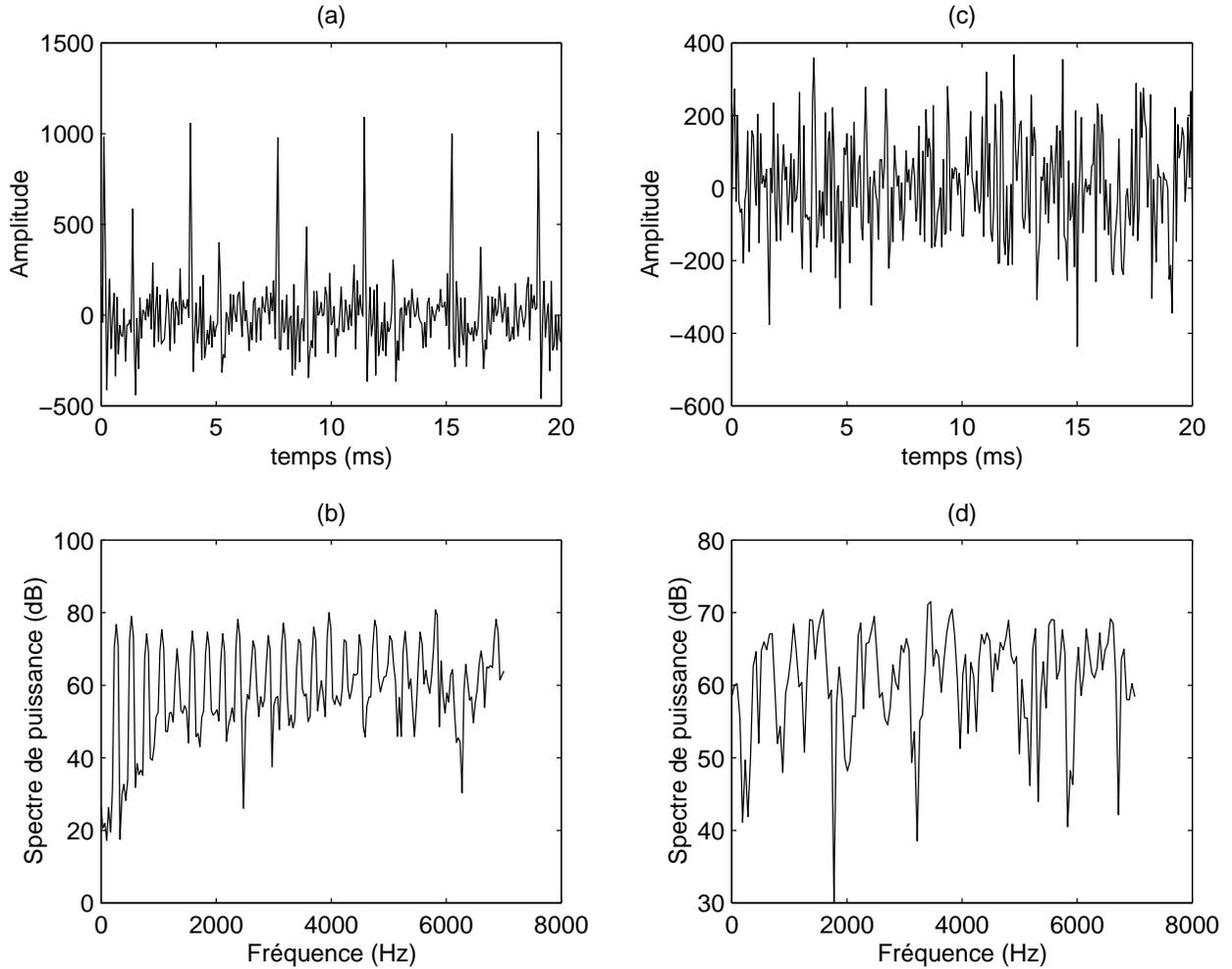}
\par\end{centering}

\caption[Excitation pour les phonèmes voisés et non voisés]{Illustration des différences entre l'excitation pour un phonème voisé et un phonème non voisé. (a) L'excitation du phonème voisé /a/ (voyelle) dans le domaine temporel et (b) fréquentiel. (c) L'excitation du phonème non voisé /s/ (fricative) dans le domaine temporel et (d) fréquentiel.}\label{fig_excitation}
\end{figure}

Il y a relativement peu de différence entre les excitations des différents
phonèmes. La principale différence se situe entre les phonèmes voisés
et non voisés. En effet, tel qu'illustré à la figure \ref{fig_excitation},
l'excitation pour les phonèmes voisés (/e/) peut être approximée par
un train d'impulsions et a donc un spectre comportant des harmoniques
espacées de façon régulière. L'excitation pour une fricative (/s/),
quant à elle, ressemble plutôt à du bruit blanc et son spectre ne
comporte donc pas d'harmoniques.

\section{Analyse LPC et domaine de représentation}

La prédiction linéaire est un outil indispensable dans le domaine
du traitement et du codage de la parole. En prédisant un échantillon
$x(n)$ à partir d'une combinaison linéaire des échantillons $x(n-i)$
passés, elle permet d'estimer la corrélation à court terme du signal.
Comme cette corrélation à court terme du signal de parole est l'effet
du filtrage de l'excitation par le conduit vocal, la prédiction linéaire
permet d'estimer les caractéristiques du conduit vocal et par conséquent,
l'enveloppe spectrale. Un fois le filtre du conduit vocal identifié,
il est alors facile d'obtenir l'excitation glottale. Ainsi, soient
$a_{i}$ les coefficients de prédiction linéaire, le signal de parole
$x(n)$ est repésenté par:
\begin{equation}
x(n)=\sum_{i=1}^{N}a_{i}x(n-i)+r(n)\label{eq_prediction}
\end{equation}

où $r(n)$ est appelé \og résidu de prédiction \fg{} et représente
le signal d'excitation. Si on considère les coefficients $a_{i}$
comme représentant un filtre tout-zéros
\begin{equation}
A(z)=1-\sum_{i=1}^{N}a_{i}z^{-i}\label{eq_prediction_tout_zeros}
\end{equation}
le signal d'excitation devient alors
\begin{equation}
R(z)=A(z)X(z)\label{eq_LPC_residue}
\end{equation}

Inversement, on peut reconstruire le signal de parole à partir de
l'excitation et d'un filtre tout-pôles par
\begin{equation}
X(z)=\frac{1}{A(z)}R(z)\label{eq_LPC_synthese}
\end{equation}

L'analyse LPC permet donc de décomposer un signal de parole en son
excitation et son enveloppe spectrale et permet de refaire la synthèse
du signal original par la suite. On référera à $A(z)$ comme étant
le filtre d'analyse et à $\frac{1}{A(z)}$ comme étant le filtre de
synthèse.

\subsection{Estimation des coefficients de prédiction}

Avant d'effectuer l'analyse LPC, on multiplie généralement le signal
$x(n)$ par une fenêtre de Hanning $w(n)$ de longueur $L$ pour obtenir
un signal $x_{w}(n)$, soit
\begin{equation}
x_{w}(n)=w(n)x(n)\label{eq_fenetrage}
\end{equation}

La fenêtre de Hanning est définie par:
\begin{equation}
w(n)=\left\{ \begin{array}{ll}
0,5-0,5\cos\left(\frac{2\pi n}{L-1}\right), & \quad0\leq n<L\\
0, & \quad sinon
\end{array}\right.\label{eq_hanning}
\end{equation}

Le critère d'erreur utilisé pour estimer les coefficients de prédiction
est l'erreur quadratique entre le signal $x_{w}(n)$ et sa prédiction
linéaire, soit
\begin{equation}
E=\sum_{n=-\infty}^{\infty}\left[x_{w}(n)-\sum_{i=1}^{N}a_{i}x_{w}(n-i)\right]^{2}\label{eq_LPC_erreur}
\end{equation}

On obtient les valeurs de $a_{k}$ optimales en minimisant le critère
d'erreur $E$ de sorte que
\begin{equation}
\frac{\partial E}{\partial a_{i}}=0,\qquad i=1,2,\ldots,N\label{eq_LPC_erreur_dE}
\end{equation}
d'où on tire que
\begin{equation}
\sum_{n=-\infty}^{\infty}x_{w}(n)x_{w}(n-k)=\sum_{i=1}^{N}a_{i}\sum_{n=-\infty}^{\infty}x_{w}(n)x_{w}(n-i),\qquad k=1,2,\ldots,N\label{eq_LPC_system}
\end{equation}

Soit $R(m)$ l'autocorrélation du signal fenêtré $x_{w}(n)$ donnée
par
\begin{equation}
R(m)=\sum_{n=i}^{L-1}x_{w}(n)x_{w}(n-m)\label{eq_autocorrelation}
\end{equation}
on obtient que 
\begin{equation}
\left[\begin{array}{cccc}
R(0) & R(1) & \cdots & R(N-1)\\
R(1) & R(0) & \cdots & R(N-2)\\
\vdots & \vdots & \ddots & \vdots\\
R(N-1) & R(N-2) & \cdots & R(0)
\end{array}\right]\left[\begin{array}{c}
a_{1}\\
a_{2}\\
\vdots\\
a_{N}
\end{array}\right]=\left[\begin{array}{c}
R(1)\\
R(2)\\
\vdots\\
R(N)
\end{array}\right]\label{eq_systeme_toeplitz}
\end{equation}

Comme la matrice en \ref{eq_systeme_toeplitz} est Toeplitz-Hermitienne,
le système peut être résolu efficacement par la récursion de Levinson-Durbin\cite{levinson_durbin}
en un temps $\mathcal{O}\left(N^{2}\right)$. De plus, il peut être
démontré que tous les pôles du filtre de prédiction se trouvent à
l'intérieur de cercle unitaire dans le plan $z$, ce qui assure la
stabilité du filtre tout-pôles\cite{LPC_stabilite}.

\subsection{Conditionnement de l'analyse\label{sec_conditionnement_LPC}}

L'analyse LPC est une opération qui peut être sensible à certaines
caractéristiques du signal. Cette sensibilité vient de l'inversion
de la matrice des autocorrélations en \ref{eq_systeme_toeplitz}.
Il est donc nécessaire de s'assurer que cette matrice soit toujours
bien conditionnée, peu importe le signal d'entrée de l'analyse LPC.

Comme dans un signal de parole, l'énergie aux hautes fréquences est
généralement beaucoup plus faible que celle aux basses fréquences,
il se peut que la matrice des autocorrélations soit mal conditionnée\cite{pre_accent}.
Afin de remédier à la situation, on applique au signal une pré-accentuation
des hautes fréquences, généralement en filtrant le signal avec le
filtre passe-haut
\begin{equation}
H(z)=1-\alpha z^{-1}\label{eq_LPC_preaccent}
\end{equation}

Afin d'aider au conditionnement de la matrice des autocorrélations,
on peut légèrement augmenter les valeurs sur sa diagonale, soit $R(0)$.
Ceci est exactement équivalent à l'ajout de bruit blanc au signal
original, puisque l'autocorrélation du bruit blanc vaut $R(m)=\delta_{m,0}$.
Le fait de multiplier le coefficient $R(0)$ par $\alpha$ correspond
à un rapport signal sur bruit de
\begin{equation}
RSB=10\log_{10}\left(\frac{1}{\alpha-1}\right)\,dB\label{eq_LPC_noise_floor}
\end{equation}
Ainsi, la valeur $\alpha=1,0001$ correspond à un \og plancher de
bruit \fg{} de $40\,dB$. Ce plancher de bruit aura aussi des avantages
lors de la transformation vers le domaine cepstral.

Enfin, lorsque le conduit vocal comporte une résonance importante,
les pôles du filtre de prédiction se trouvent très près du cercle
unitaire dans le plan $z$. Dans ces cas, les erreurs d'arrondi lors
des calculs ou la quantification des coefficients LPC peut faire en
sorte que le filtre tout-pôles devienne instable. De plus, lorsque
un ou plusieurs pôles du filtre LPC se trouvent près du cercle unitaire
dans le plan $z$, il devient plus difficile de traiter ce filtre
par son enveloppe spectrale (section \ref{sec_LPC_a_enveloppe}).

Une des façons de remédier au problème est de ramener les pôles du
filtre $\frac{1}{A(z)}$ vers l'origine par un facteur $\gamma$,
ce qui correspond à utiliser le filtre $\frac{1}{A(\gamma z)}$. Malheureusement,
cette technique, appelée \emph{bandwidth expansion}\cite{LPC_stabilite},
a l'inconvénient de modifier aussi la position de tous les autres
pôles. Pour cette raison, on utilise plutôt une technique appelée
\emph{lag windowing}\cite{lag_windowing}. Cette méthode consiste
à fenêtrer la valeur des autocorrélations $R(m)$ avant de résoudre
le système en \ref{eq_systeme_toeplitz}. Comme le spectre de puissance
d'un signal n'est autre que la transformée de Fourier de l'autocorrélation,
un fenêtrage de l'autocorrélation est équivalent à un filtrage (un
lissage) du spectre de puissance. On peut utiliser une fenêtre gaussienne
\begin{equation}
w_{a}(m)=e^{-\left(\beta m\right)^{2}}\label{eq_lag_window}
\end{equation}
où le paramètre $\beta$ contrôle la largeur de bande des résonances.

\subsection{Coefficients de prédiction et enveloppe fréquentielle\label{sec_LPC_a_enveloppe}}

Lors du traitement, il est parfois désirable de calculer l'enveloppe
spectrale du signal à partir des coefficients de prédiction $a_{i}$.
Ceci correspond à calculer le spectre de la réponse impulsionnelle
du filtre de synthèse $\frac{1}{A(z)}$. Comme cette réponse impulsionnelle
est théoriquement infinie et peut, en pratique, être très longue si
le filtre comporte un pôle près du cercle unitaire, on calculera plutôt
le spectre de la réponse du filtre d'analyse $A(z)$, puis on inversera
ce spectre. Ainsi, le spectre $S_{a}(k)$ du filtre d'analyse est
\begin{equation}
S_{a}(k)=\left|\sum_{i=0}^{N}a_{i}e^{-2\pi\jmath ik/N}\right|^{2}\label{eq_spectre_analyse}
\end{equation}

Le spectre de puissance du filtre de synthèse est donc
\begin{equation}
S_{s}(k)=\frac{1}{S_{a}(k)}\label{eq_spectre_synthese}
\end{equation}

Il faut noter que le nombre de points utilisé pour la transformée
de Fourier est d'une grande importance. En effet, si le nombre de
points est trop faible, les résonances du filtre LPC ne seront pas
\og vues \fg{} lors de la transformation. La longueur de la transformée
est donc fonction du paramètre $\beta$ utilisé lors du \emph{lag
windowing}. En effet, la \og distance \fg{} entre deux points du
spectre doit être au moins égale à la largeur minimale des résonances
du filtre LPC.

Enfin, il est possible d'effectuer l'opération inverse, soit de retrouver
les coefficients de prédiction $a_{i}$ à partir du spectre de puissance
du filtre de synthèse. En effet, étant donné $S_{s}(k)$, on peut
trouver les autocorrélations $R(m)$ par la transformée de Fourier
inverse, soit

\begin{equation}
R(m)=\sum_{k=0}^{N}S_{s}(k)e^{2\pi\jmath km/N}\label{eq_spectre_autocorr}
\end{equation}

Une fois les autocorrélations trouvées, on peut retrouver les coefficients
de prédiction par la récursion de Levinson-Durbin.

\subsection{Le domaine cepstral}

Le domaine cepstral représente de façon harmonique le logarithme du
spectre d'un signal de parole. Plus précisément, le cepstre $C_{s}(T)$
de l'enveloppe spectrale $S_{s}(k)$ s'exprime comme suit: 
\begin{equation}
C_{s}(T)=\mathcal{IFFT}\left\{ \log\left(S_{s}(k)\right)\right\} \label{eq_cepstre}
\end{equation}

Ce domaine de représentation comporte plusieurs avantages. D'abord,
le fait de traiter l'enveloppe spectrale dans le domaine logarithmique
se rapproche plus de la sensibilité de l'oreille. Ensuite, la transformée
de Fourier inverse a pour effet de décorréler les valeurs du spectre
et d'en \og concentrer \fg{} l'énergie dans les quelques premiers
coefficients cepstraux. Typiquement, de 8 à 12 coefficients cepstraux
sont suffisants pour représenter adéquatement l'enveloppe spectrale
dans la bande téléphonique. Il faut noter que $C_{s}(T)$ représente
le cepstre de la réponse temporelle du filtre de synthèse, et non
le cepstre du signal de parole lui-même.

Il faut noter que la terminologie utilisée pour le domaine cepstral
est obtenue en inversant le mot correspondant pour le domaine spectrale.
Ainsi, \emph{spectre}, \emph{fréquence} et \emph{filtrer} deviennent
respectivement \emph{cepstre}, \emph{quéfrence} et \emph{liftrer}
pour le domaine cepstral.

\section{Analyse du pitch\label{sec_analyse_pitch}}

Le signal de parole comporte des corrélations, autant à court terme
qu'à long terme. Les corrélations à court terme forment l'enveloppe
spectrale et sont produites par les résonances du conduit vocal. Les
corrélations à long terme quant à elles, représentent le pitch de
la voix et sont le résultat des impulsions glottales. Alors que l'analyse
LPC modélise les corrélations à court terme du signal de parole, on
utilise un filtre prédicteur long terme pour modéliser le pitch comme
\begin{equation}
x(n)\approx\beta x(n-T)\label{eq_approx_pitch}
\end{equation}
où $T$ est la période du pitch (la durée entre les impulsions glottales)
et $\beta$ est appelé \og gain de pitch \fg{}. La fonction de transfert
du filtre prédicteur long-terme est donc:
\begin{equation}
H(z)=1-\beta z^{-T}\label{eq_pitch_predict}
\end{equation}

Dans l'intervalle $\left[M\,N\right]$, les paramètres du pitch peuvent
être trouvés en minimisant la fonction d'erreur: 
\begin{equation}
E=\sum_{i=M}^{N}\left[x(i)-\beta x(i-T)\right]^{2}\label{eq_pitch_error}
\end{equation}
En posant $\frac{\partial E}{\partial\beta}=0$, on trouve: 
\begin{eqnarray}
0 & = & \sum_{i=M}^{N}\left[x(i)-\beta x(i-T\right]x(i-T)\label{eq_pitch_beta1}\\
\sum_{i=M}^{N}x(i)x(i-T) & = & \beta\sum_{i=M}^{N}x(i)x(i-T)\label{eq_pitch_beta2}\\
\beta & = & \frac{\sum_{i=M}^{N}x(i)x(i-T)}{\sum_{i=M}^{N}x(i-T)^{2}}\label{eq_pitch_beta3}
\end{eqnarray}

Ainsi, en remplaçant \ref{eq_pitch_beta3} dans \ref{eq_pitch_error},
on obtient:
\begin{eqnarray}
E & = & \sum_{i=M}^{N}x(i)x(i)-2\beta x(i)x(i-T)+\beta^{2}x(i-T)x(i-T)\label{eq_pitch_T1}\\
E & = & \sum_{i=M}^{N}x(i)x(i)-\frac{\left(\sum_{i=M}^{N}x(i)x(i-T)\right)^{2}}{\sum_{i=M}^{N}x(i-T)^{2}}\label{eq_pitch_T2}\\
T & = & \max_{T}\,\left[\frac{\left(\sum_{i=M}^{N}x(i)x(i-T)\right)^{2}}{\sum_{i=M}^{N}x(i-T)^{2}}\right]\label{eq_pitch_T3}
\end{eqnarray}

Pour les phonèmes voisés, le gain de pitch $\beta$ est souvent près
de $1$, alors que pour les phonèmes non voisés, $\beta$ est plus
près de $0$. Bien que $\beta$ soit généralement compris entre $0$
et $1$, il peut arriver que dans certaines circonstances, il en soit
autrement.

\chapter{Modèle d'extension proposé\label{sec_overview}}

Dans le projet d'extension de la bande téléphonique à la bande AM,
on note deux principales parties, soit l'extension des hautes fréquences
et l'extension des basses fréquences. Ces deux parties ont relativement
peu en commun l'une avec l'autre. De cette façon, il est possible
de choisir pour chaque bande, le modèle le plus approprié. Cette façon
de faire comporte aussi des avantages pour le design et le débogage
de ces parties. En effet, il est ainsi possible d'isoler les artefacts
causés par l'extension des deux bandes de fréquence. 

\begin{figure}[!t]
\begin{centering}
\includegraphics[width=1\columnwidth]{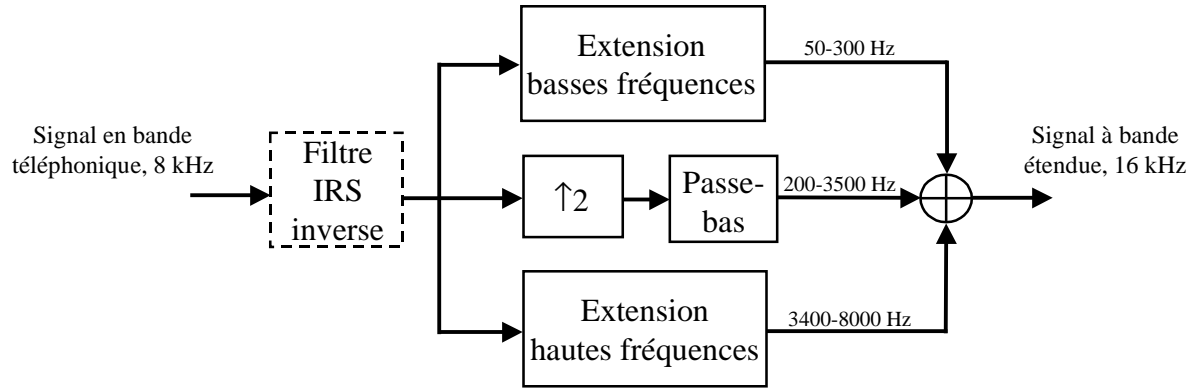}
\par\end{centering}

\caption[Vue d'ensemble du système]{Vue d'ensemble du système d'extension de la bande.}\label{fig_system_overview}
\end{figure}

Le signal large bande reconstruit sera constitué des trois parties:
\begin{itemize}
\item la bande téléphonique originale, de $200\,\mathrm{Hz}$ à $3500\,\mathrm{Hz}$;
\item la bande basse reconstruite, de $50\,\mathrm{Hz}$ à $200\,\mathrm{Hz}$;
\item la bande haute reconstruite, de $3500\,\mathrm{Hz}$ à $7000\,\mathrm{Hz}$.
\end{itemize}
Comme le signal original dans la bande téléphonique est échantillonné
à $8\,\mathrm{kHz}$, il sera d'abord nécessaire de le sur-échantillonner
par un facteur 2, puis de filtrer passe-bas afin de ne garder que
la partie sous $3500\,\mathrm{kHz}$, évitant ainsi le repliement
spectral causé par le sur-échantillonnage. Une fois la bande téléphonique
échantillonnée à $16\,\mathrm{kHz}$, on pourra la combiner aux deux
bandes étendues pour obtenir le signal large bande reconstruit à $16\,\mathrm{kHz}$.
Le processus est illustré à la figure \ref{fig_system_overview}.
Il faut noter que la sortie des blocs \og extension basses fréquences \fg{}
et \og extension hautes fréquences \fg{} est échantillonnée à $16\:\mathrm{kHz}$.
Le bloc \og filtre IRS modifié inverse \fg{} est optionnel et décrit
à la section \ref{sec_IRM_INV}.

Étant donné la diversité du public visé pour ce projet, le système
conçu devra pouvoir fonctionner pour différents locuteurs parlant
différentes langues. Pour cette raison, les différentes étapes d'entraînement
devront utiliser une base de donnée multilingue et multilocuteur.

\section{Points principaux du système d'extension}

Le sous-système d'extension des hautes fréquences est basé sur le
modèle filtre-excitation, permettant de reconstruire séparément l'enveloppe
spectrale (filtre) et l'excitation. L'extension de l'excitation s'effectue
dans le domaine temporel par un système non linéaire. L'enveloppe
spectrale est reconstruite par hétéro-association entre l'enveloppe
spectrale de la bande téléphonique et l'enveloppe spectrale de la
bande haute.

Le sous-système d'extension des basses fréquences utilise plutôt le
modèle sinusoïdal pour reconstruire les deux premières harmoniques
du pitch. Leur amplitude est, elle aussi, estimée par hétéro-association
à partir de paramètres calculés sur la bande téléphonique.

\section{Inversion du IRS modifié\label{sec_IRM_INV}}

\begin{figure}[!t]
\begin{centering}
\includegraphics[width=1\columnwidth]{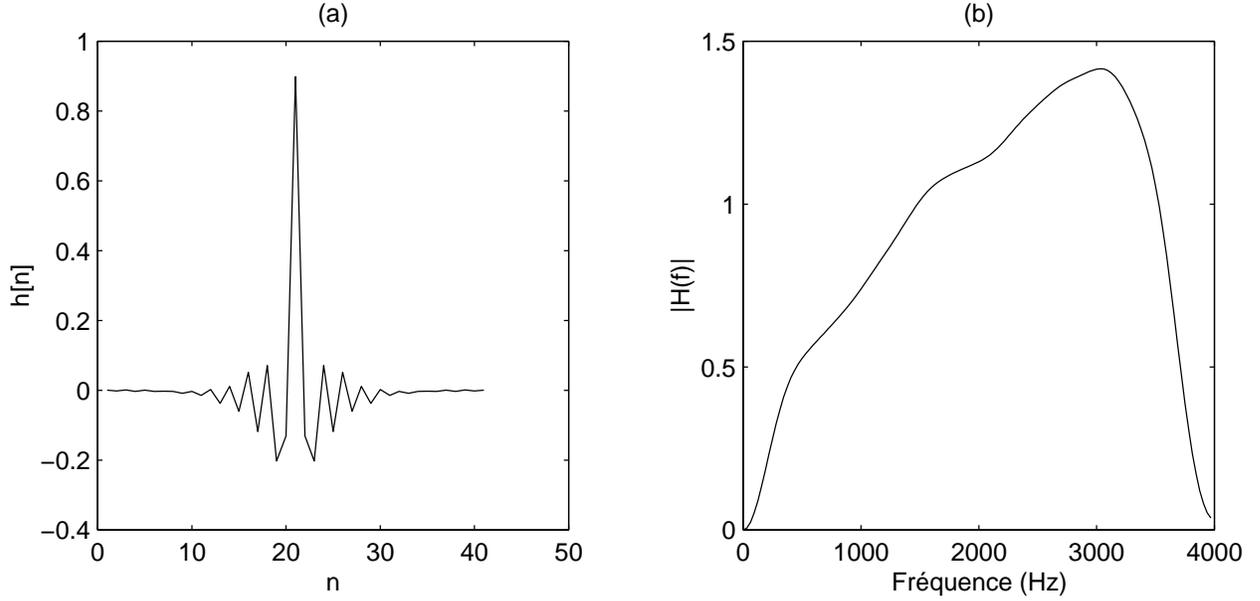}
\par\end{centering}

\caption[Réponse du filtre IRS modifié]{Réponse (a) temporelle et (b) fréquentielle du filtre IRS modifié}\label{fig_IRM}
\end{figure}

Dans la majorité des applications téléphoniques, le signal échantillonné
à $8\,\mathrm{kHz}$ est filtré de manière à améliorer l'intelligibilité,
ainsi que la qualité perçue par l'utilisateur. Le filtre FIR utilisé
est appelé \og filtre IRS modifié \fg{} et sa réponse temporelle
et fréquentielle est montrée à la figure \ref{fig_IRM}. On considère
généralement que la bande passante de ce filtre est entre $200\,\mathrm{Hz}$
à $3500\,\mathrm{Hz}$ et le gain maximal se trouve aux environs de
$3000\,\mathrm{Hz}$. Aussi, comme la réponse temporelle est symétrique,
le filtre comporte une réponse de phase linéaire.

Avant de pouvoir faire l'extension de la bande, il est d'abord nécessaire
de s'assurer que la réponse dans la bande téléphonique soit constante,
telle qu'elle était dans le signal original. Pour cela, il est nécessaire
d'utiliser un filtre dont la réponse fréquentielle est l'inverse de
celle du filtre IRS modifié. On considère un filtre FIR à phase linéaire
dont la réponse fréquentielle est
\begin{equation}
\left|H(\omega)\right|=a_{0}+2a_{1}\cos\,\omega+2a_{2}\cos\,2\omega+\ldots+2a_{N}\cos\,N\omega\label{fig_FIR_LIN}
\end{equation}

\begin{figure}[!t]
\begin{centering}
\includegraphics[width=1\columnwidth]{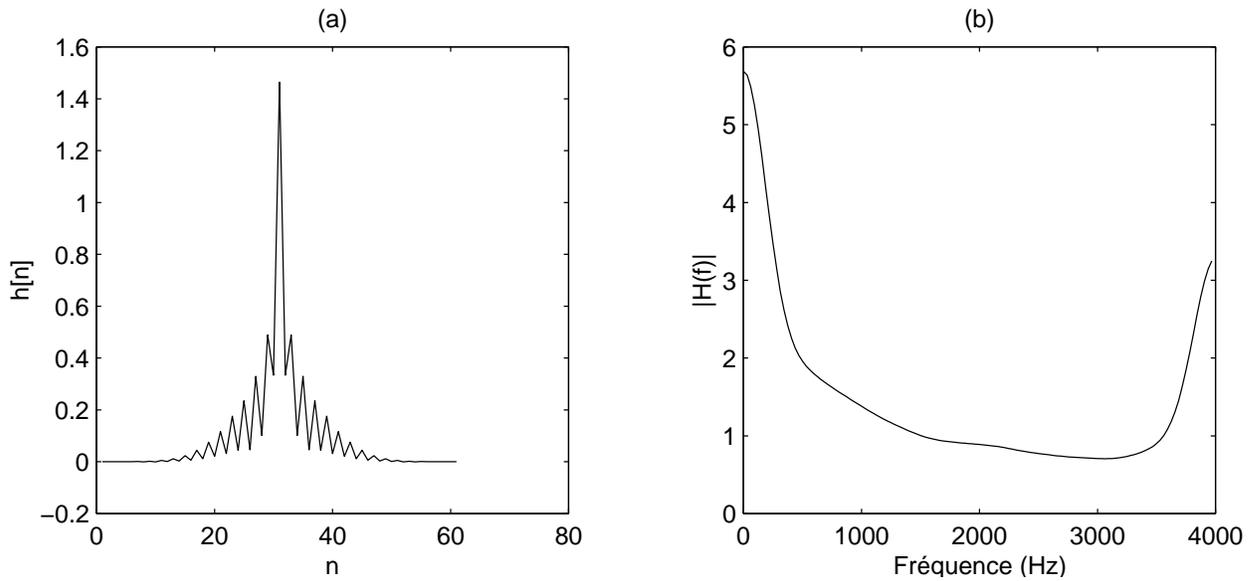}
\par\end{centering}

\caption[Réponse du filtre inverse]{Réponse (a) temporelle et (b) fréquentielle de l'inverse du filtre IRS modifié.}\label{fig_IIRM}
\end{figure}

\begin{figure}[!t]
\begin{centering}
\includegraphics[width=1\columnwidth]{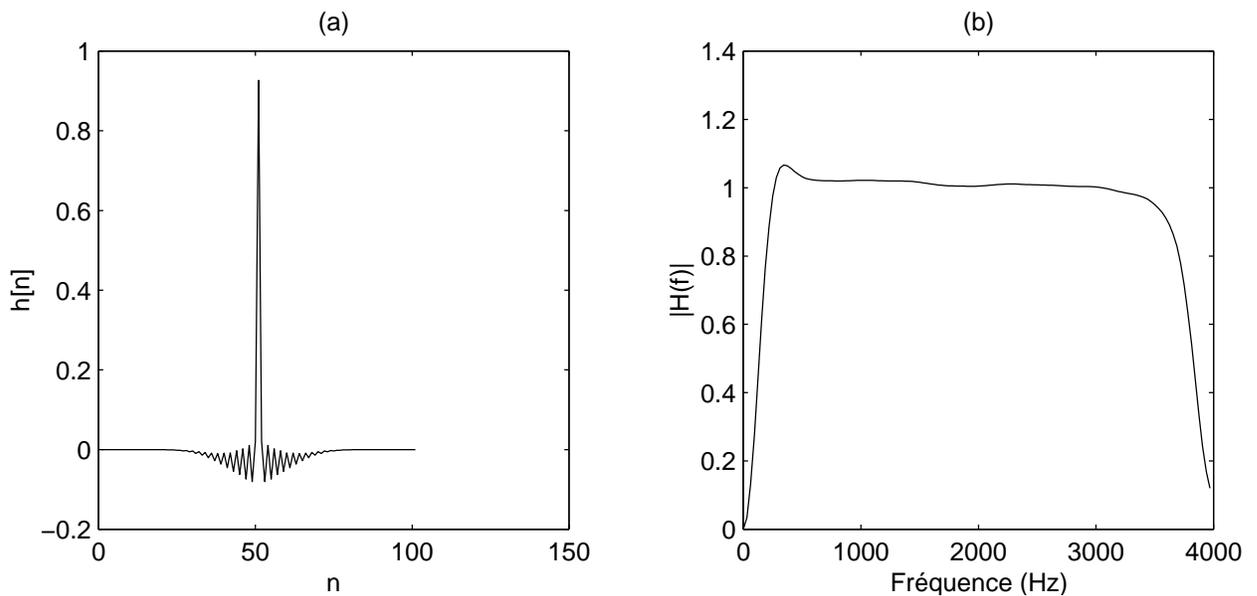}
\par\end{centering}

\caption[Réponse combinée du filtre IRS modifié et de son filtre inverse]{Réponse combinée du filtre IRS modifié et de son filtre inverse. (a) Réponse temporelle. (b) Réponse fréquentielle.}\label{fig_IRM_CONV}
\end{figure}

Aussi, soit $G(\omega)$ la réponse fréquentielle du filtre IRS modifié,
on cherche les coefficients $a_{k}$ qui font que
\begin{equation}
\left|G(\omega)\right|\left|H(\omega)\right|\approx1,\qquad\mathrm{pour}\,\omega_{1}<\omega<\omega_{2}\label{fig_IRM_INV}
\end{equation}
 où $\omega_{1}=2\pi\frac{200\,\mathrm{Hz}}{8\,\mathrm{kHz}}=0,16\,rad/s$
et $\omega_{2}=2\pi\frac{3500\,\mathrm{Hz}}{8\,\mathrm{kHz}}=2,75\,rad/s$.
Ainsi, si on minimise, dans l'intervalle $\omega_{1}<\omega<\omega_{2}$,
l'erreur quadratique
\begin{equation}
E=\left[\frac{1}{\left|G(\omega)\right|}-a_{0}+2a_{1}\cos\,\omega+2a_{2}\cos\,2\omega+\ldots+2a_{N}\cos\,N\omega\right]^{2}\label{fig_IRM_MMSE}
\end{equation}
on obtient un filtre FIR à phase linéaire qui, lorsque convolué avec
le filtre IRS modifié $G(\omega)$, produit une réponse constante
entre $200\,\mathrm{Hz}$ et $3500\,\mathrm{Hz}$. Ainsi, si on choisit
$N=30$, on obtient le filtre inverse dont les réponses temporelle
et fréquentielle sont montrés à la figure \ref{fig_IIRM}. En convoluant
ce filtre inverse avec le filtre IRS modifié, on obtient la réponse
combinée montrée à la figure \ref{fig_IRM_CONV}. On constate ainsi,
que la réponse est en effet constante entre $200\,\mathrm{Hz}$ et
$3500\,\mathrm{Hz}$; le but est donc atteint.


\chapter{Extension des hautes fréquences\label{sec_ext_HF}}

\section{Modèle utilisé}

La plupart des articles\cite{Epps,Enbom,Chan,Ming,Jax} traitant de
l'extension spectrale dans la bande haute utilisent le modèle \og filtre-excitation \fg{}
de la parole et c'est ce modèle qui a été retenu pour les présents
travaux. Selon ce modèle, décrit à la section \ref{sec_signal_et_modelisation},
le signal de parole est représenté par un filtre prédicteur à court
terme excité par un signal d'excitation \og blanc \fg{}, c'est-à-dire
dont le spectre est globalement plat. Le filtre prédicteur à court
terme est obtenu à partir d'une analyse LPC sur le signal de parole
et représente l'enveloppe spectrale du signal. 

\begin{figure}[!t]
\begin{centering}
\includegraphics[width=1\columnwidth]{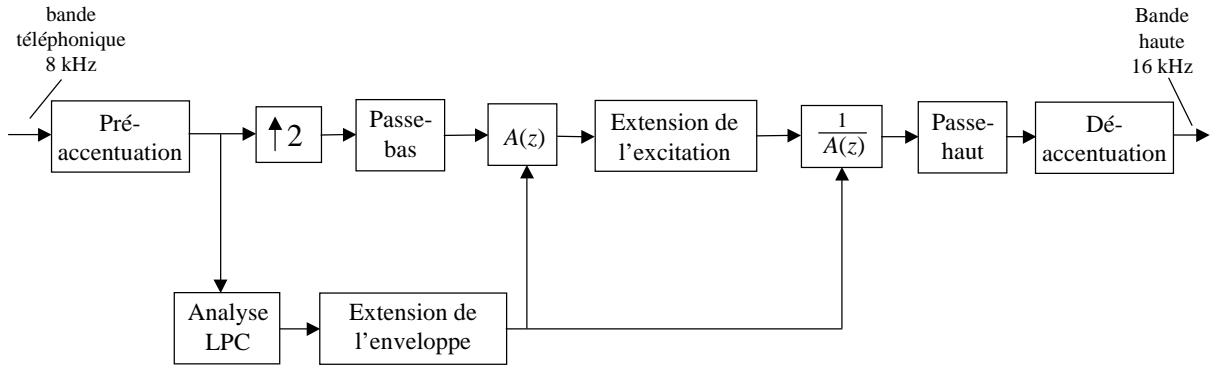}
\par\end{centering}

\caption[Modèle d'extension des hautes fréquences]{Modèle filtre-excitation pour l'extension des hautes fréquences}\label{fig_hf_modele}
\end{figure}

Lors de l'extension de la bande haute, l'excitation et le filtre prédicteur
sont reconstruits séparément. Plutôt que de calculer l'excitation
en filtrant le signal échantillonné à $8\,\mathrm{kHz}$ par le filtre
LPC calculé sur la bande téléphonique, il est préférable de sur-échantillonner
ce signal à $16\,\mathrm{kHz}$ et de calculer l'excitation à partir
des nouveaux coefficients LPC large bande\cite{Jax}. En plus de simplifier
les calculs, ceci permet de réduire certains artefacts qui résulteraient
des légères différences entre les filtre d'analyse et de synthèse.
Le modèle d'extension pour les hautes fréquences est illustré à la
figure \ref{fig_hf_modele}.

Bien que le modèle filtre-excitation soit le plus utilisé, il faut
noter que d'autres travaux sur l'extension de la bande haute sont
plutôt basés sur le modèle sinusoïdal\cite{Epps2}.

\section{Extension de l'excitation}

Le problème de l'extension spectrale du signal d'excitation est connu
depuis les années 1970. En effet l'extension de l'excitation a d'abord
été utilisée dans les codeurs de parole de type RELP au cours des
années 1970. Cependant, pour le présent projet, l'extension se fera
jusqu'à $8000\,\mathrm{Hz}$, plutôt que $4000\,\mathrm{Hz}$.

\subsection{Repliement spectral et non linéarité\label{eq_excitation_abs}}

La méthode la plus utilisée pour étendre la bande de l'excitation
est le repliement spectral\cite{Berouti}. Le signal d'excitation
en bande téléphonique $r_{bt}(n)$ est transformé en un signal d'excitation
large bande $r_{lb}(n)$
\begin{equation}
r_{lb}(n)=\left\{ \begin{array}{cl}
2r_{bt}(n)\qquad & n\,\mathrm{pair}\\
0\qquad & n\,\mathrm{impair}.
\end{array}\right.\label{eq_folding}
\end{equation}

Le spectre résultant est alors une image miroir du spectre dans la
bande basse. Il est cependant connu que cette méthode, en plus de
causer des discontinuités dans le spectre, produit un son trop harmonique
aux hautes fréquences. De plus, cette méthode ne fonctionne bien que
si l'on effectue l'extension de la bande $0-4000\:\mathrm{Hz}$ à
la bande $0-8000\:\mathrm{Hz}$. Dans le cas où l'on fait l'extension
à partie de la bande téléphonique en utilisant un repliement spectral
par deux, on obtient un \og trou \fg{} dans la bande $3500\:\mathrm{Hz}$
à $4500\:\mathrm{Hz}$\cite{Jax}.

\begin{figure}[!t]
\begin{centering}
\includegraphics[width=1\columnwidth]{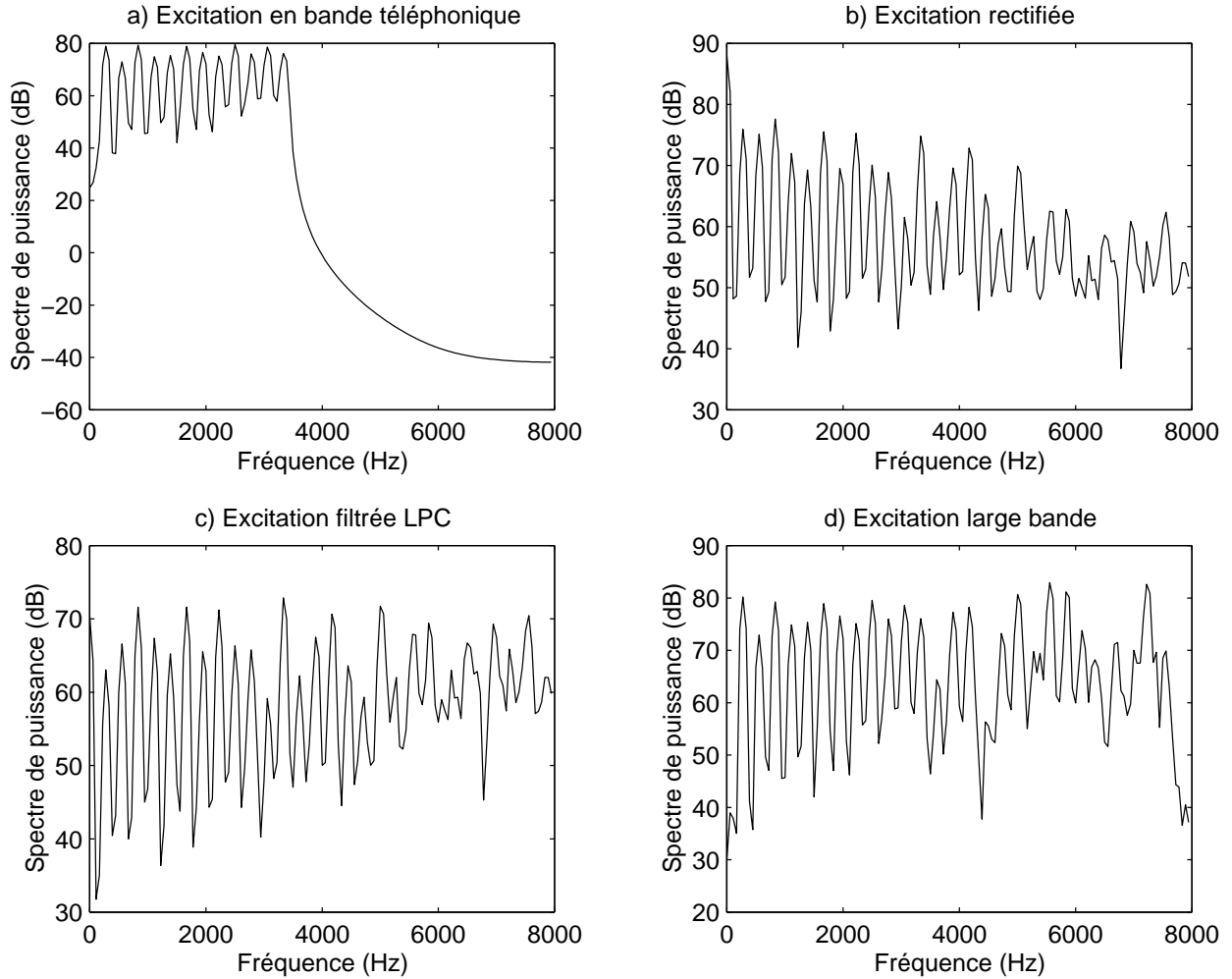}
\par\end{centering}

\caption[Extension de l'excitation]{Spectre du signal d'excitation à différentes étapes du processus d'extension. (a) Excitation filtrée passe-bas à $3500\, \mathrm{Hz}$. (b) Excitation redressée. (c) Excitation redressée, filtrée par le filtre LPC. (d) Excitation originale large bande.}\label{fig_ext_excitation}
\end{figure}

Une méthode alternative consiste à utiliser une distortion non linéaire
dans le domaine temporel, produisant ainsi des harmoniques dans la
bande haute. La fonction \og valeur absolue \fg{} (redresseur pleine-onde)
constitue un bon choix pour cette fonction non linéaire\cite{Berouti},
car elle est très peu complexe à calculer et affecte peu l'amplitude
du signal (contrairement à la fonction $x^{2}$). Comme le spectre
du signal d'excitation après redressement (fig. \ref{fig_ext_excitation}b)
n'est plus plat comme il doit l'être (voir section \ref{sec_caract_parole}),
il est nécessaire de le filtrer de manière à le rendre plat (fig.
\ref{fig_ext_excitation}c). Ceci peut être accompli en utilisant
un filtre LPC\cite{Weinstein,Valin}, après quoi il est nécessaire
de renormaliser le signal de telle sorte que l'énergie dans la bande
téléphonique demeure la même que pour l'excitation originale. 

On voit que le signal résultant est très semblable au signal d'excitation
large bande original (fig. \ref{fig_ext_excitation}d). Il faut noter
que les distorsions produites aux fréquences inférieures à $3500\mathrm{Hz}$
n'auront pas d'effet sur le signal reconstruit, puisque le résultat
de l'extension aux hautes fréquences est filtré passe-haut (fig. \ref{fig_hf_modele})
avant d'être combiné au signal original en bande téléphonique (fig.
\ref{fig_system_overview}).

\subsection{Application au codage large bande}

En plus de l'application au présent problème de l'extension spectrale,
la méthode d'extension de l'excitation présentée à la section \ref{eq_excitation_abs}
peut avoir des applications pour le codage de la parole large bande.
En effet, certaines expériences montrent que lorsque l'on utilise
le filtre LPC calculé sur le signal large bande et que l'on effectue
l'extension de l'excitation entre $3500\,\mathrm{Hz}$ et $7000\,\mathrm{Hz}$,
le signal résultant est presque impossible à distinguer de l'original. 

\begin{figure}[!t]
\begin{centering}
\includegraphics[width=1\columnwidth]{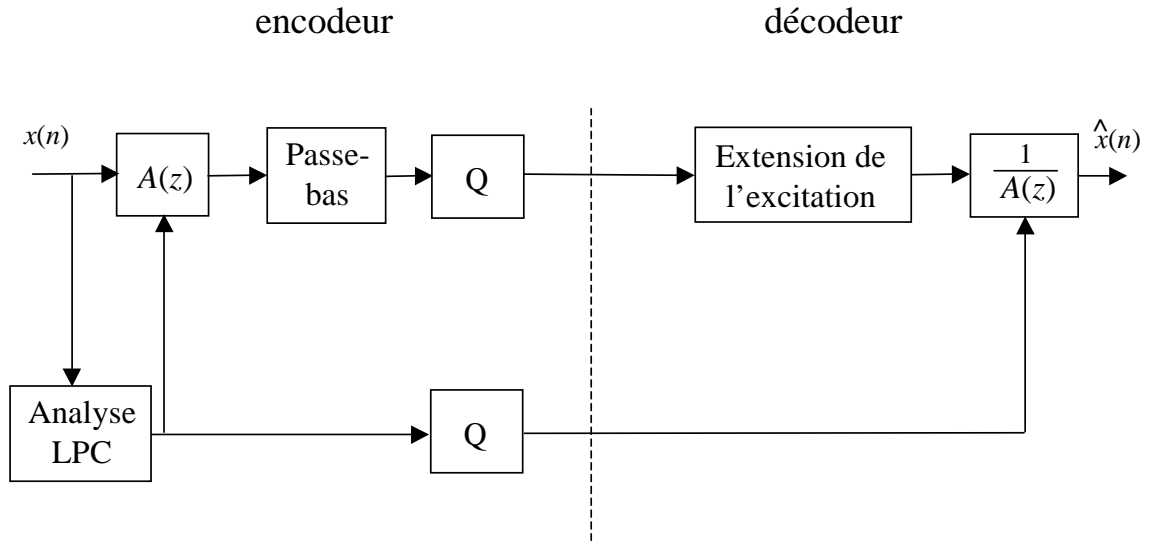}
\par\end{centering}

\caption[Application au codage de parole]{Codeur large bande utilisant l'extension de l'excitation. L'enveloppe spectrale est codée dans la bande $0-8000\, \mathrm{Hz}$, l'excitation est codée dans la bande $0-4000\, \mathrm{Hz}$.}\label{fig_exc_coding}
\end{figure}

Ceci rend la technique intéressante pour le codage large bande, car
il serait alors possible de coder l'excitation jusqu'à $3500\,\mathrm{Hz}$
seulement, réduisant ainsi l'information transmise. La figure \ref{fig_exc_coding}
illustre le mode de fonctionnement d'un codeur qui utiliserait l'extension
de l'excitation afin de ne coder que la bande basse de l'excitation,
diminuant ainsi le débit tout en maintenant la qualité de la parole.

\begin{figure}[!t]
\begin{centering}
\includegraphics[width=1\columnwidth]{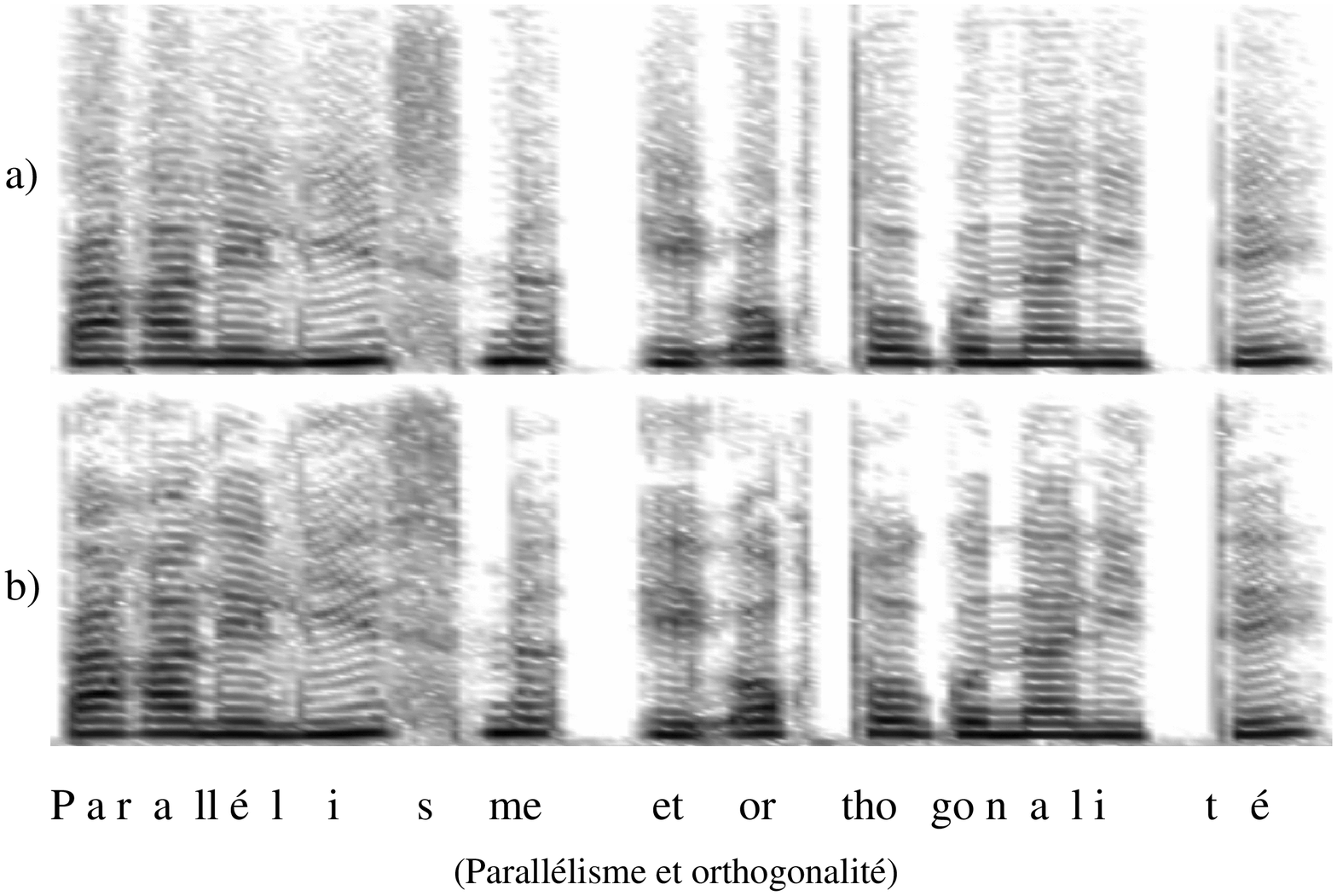}
\par\end{centering}

\caption[Codage par extension de l'excitation]{Résultat de l'extension de l'excitation lorsque l'enveloppe spectrale est préservée. Le spectrogramme du signal de parole reconstruit (a) est comparé à celui du signal original (b).}\label{fig_spectrogramme_excitation}
\end{figure}

La constatation du haut niveau de qualité obtenue pour l'extension
de l'excitation avait aussi été faite pour un système utilisant le
modèle sinusoïdal\cite{Epps2}, rendant les deux modèles équivalents,
puisque la qualité du résultat dépend essentiellement de la qualité
de l'extension de l'enveloppe spectrale. Le résultat de l'extension
de l'excitation peut être observé sur le spectrogramme de la figure
\ref{fig_spectrogramme_excitation}. Tout comme lors de l'écoute,
on note très peu de différence entre le spectrogramme du signal original
et celui du signal avec extension de l'excitation. Le spectrogramme
montre aussi que la structure harmonique du signal dans les hautes
fréquences est en général bien reconstruite.

Il est important de savoir que, étant donné le modèle utilisé, le
système de codage par extension de l'excitation présenté à la figure
\ref{fig_exc_coding} ne peut fonctionner correctement que pour la
parole. Aussi, les résultats de l'extension pour un signal musical
sont plutôt mauvais. Toutefois, il est possible d'appliquer la méthode
au codage de signaux musicaux haute-fidélité échantillonnés à $44,1\:\mathrm{kHz}$
pour faire l'extension entre $11\;\!\mathrm{kHz}$ à $22\:\mathrm{kHz}$.
Comme l'oreille humaine est très peu sensible à ces fréquences, même
des signaux de musique codés ainsi conservent une bonne qualité. Des
expériences préliminaires ont montré qu'il était possible d'obtenir
une qualité sonore acceptable en n'utilisant que $345\:bits/s$ ($8\:bits/trame$,
pour des trames de $1024$ échantillons) pour quantifier l'enveloppe
spectrale haute ($11-22\:\mathrm{kHz}$).

\section{Extension de l'enveloppe spectrale}

L'extension hautes fréquences de l'enveloppe spectrale consiste à
estimer un filtre LPC large bande ($F_{e}=16\:\mathrm{kHz}$) à partir
d'un filtre LPC en bande téléphonique ($F_{e}=8\:\mathrm{kHz}$) et
de quelques paramètres additionnels, calculés aussi sur la bande téléphonique.
Ce problème général est connu sous le nom d'\emph{hétéro-association}
soit l'association entre un ensemble de formes d'entrée et un ensemble
de formes de sortie\cite{Mehrotra}.

Lors de l'extension de l'enveloppe, les coefficients LPC calculés
sur la bande téléphonique sont convertis en une enveloppe spectrale.
Cette enveloppe, ainsi que d'autres paramètres, sont utilisés comme
entrée de l'algorithme d'extension de l'enveloppe spectrale grâce
auquel, on obtient l'enveloppe hautes fréquences. Les deux bandes
sont ensuite concaténées pour obtenir une enveloppe spectrale large
bande, qui est finalement convertie en un filtre LPC large bande.
C'est ce filtre qui sera utilisé, autant pour l'analyse que pour la
synthèse.

\subsection{Mesure de distorsion spectrale}

Avant de décrire l'extension de l'enveloppe spectrale, il est nécessaire
d'introduire la notion de distorsion spectrale qui sera utilisée comme
mesure d'erreur quantitative pour cette partie. De façon générale,
pour une trame $k$ donnée, la distorsion spectrale $D_{k}$, exprimée
en $\mathrm{dB}$, entre la vraie enveloppe spectrale $A_{k}(\omega)$
et une approximation $\widetilde{A}_{k}(\omega)$ est: 
\begin{equation}
D_{k}=\sqrt{\frac{1}{\omega_{2}-\omega_{1}}\int_{\omega_{1}}^{\omega_{2}}\left[20\log_{10}\left|\frac{A_{k}\left(\omega\right)}{\widetilde{A}_{k}\left(\omega\right)}\right|\right]^{2}d\omega}\label{eq_distorsion_spectrale}
\end{equation}
où $\left[\omega_{1},\:\omega_{2}\right]$ est l'intervalle de fréquence
pour lequel on mesure la distortion spectrale. Par exemple, pour mesurer
la distortion spectrale dans le bande haute, on utilise $\omega_{1}=3500\:\mathrm{Hz}$
et $\omega_{2}=8000\:\mathrm{Hz}$.

La distorsion spectrale totale pour un ensemble de trames est la moyenne
en quadrature des distorsions spectrales individuelles, soit: 
\begin{equation}
D=\sqrt{\frac{1}{K}\sum_{k=1}^{K}\left[D_{k}\right]^{2}}\label{eq_dist_quadrature}
\end{equation}

\subsection{Représentation de l'enveloppe spectrale}

Afin de faciliter le traitement de l'enveloppe spectrale, tous les
calculs d'extension de l'enveloppe sont effectués dans le domaine
fréquentiel plutôt que dans le domaine des coefficients de prédiction
$a_{i}$. On utilise donc la transformation présentée à l'équation
\ref{eq_spectre_synthese} pour représenter le filtre de synthèse
$\frac{1}{A(z)}$. Lors du traitement, la bande complète $0-8000\:\mathrm{Hz}$
est divisée en un spectre $S_{lb}(k)$ de $64$ points\footnote{La transformée de Fourier discrète est donc calculée sur $128$ points.},
soit à intervalle de $\frac{8000\:\mathrm{Hz}}{64}=125\:\mathrm{Hz}$.
De la même manière, la bande $0-4000\:\mathrm{Hz}$ est divisée en
un spectre $S_{bt}(k)$ de $32$ points, pour conserver le même intervalle
entre les points. Pour la bande téléphonique, on n'utilise que les
points $S_{bt}(k),\;2\leq k\leq28$, soit la partie correspondant
à la bande $250-3500\:\mathrm{Hz}$. Pour la bande haute, prédite
à partir du signal en bande téléphonique, on garde les points $S_{lb}(k),\;24\leq k\leq63$,
soit la bande $3000-8000\:\mathrm{Hz}$.

Comme le spectre du filtre de synthèse est échantillonné à intervalles
de $125\:\mathrm{Hz}$, on doit s'assurer que ses résonances aient
une largeur de bande d'au moins $250\:\mathrm{Hz}$ afin de bien modéliser
l'enveloppe spectrale\footnote{Cette condition est semblable à celle imposée pour l'échantillonnage
d'un signal temporel, où le signal doit être préalablement filtré
à la fréquence de Nyquist, $F_{e}/2$.}. Ceci peut être contrôlé par le paramètre $\beta$ utilisé pour le
\emph{lag windowing} (section \ref{sec_conditionnement_LPC}). De
même, afin de réduire la dynamique de l'enveloppe spectrale, on la
considère dans le domaine logarithmique. Ceci a comme avantage de
placer le problème dans un domaine plus proche du domaine de perception
de l'oreille.

Il est souvent désirable de réduire le nombre de dimensions nécessaires
pour traiter un signal (dans ce cas, l'enveloppe spectrale). Ceci
peut être accompli en utilisant une transformation linéaire ne conservant
que les composantes principales du signal. La transformation optimale
pour un signal donné peut être obtenue par analyse en composante principale\cite{Haykin}.
Cependant, pour l'enveloppe spectrale d'un signal de parole, la transformée
en cosinus discrète (DCT) est presque optimale et il existe des algorithmes
pouvant calculer cette transformée en un temps $\mathcal{O}(n\log_{2}n)$.
De plus, la transformée en cosinus du spectre dans le domaine logarithmique
n'est autre que le cepstre réel du signal de parole.

Les coefficients $c_{k}$ de la transformée en cosinus du signal $x_{i}$
sont:
\begin{equation}
c_{k}=\left\{ \begin{array}{ccl}
\sqrt{\frac{1}{N}}\sum_{i=0}^{N-1}x_{i} & \qquad & k=0\\
\sqrt{\frac{2}{N}}\sum_{i=0}^{N-1}x_{i}\cos\frac{\pi\left(2i+1\right)k}{N} & \qquad & k\neq0
\end{array}\right.\label{eq_def_DCT}
\end{equation}

\begin{figure}[!t]
\begin{centering}
\includegraphics[width=1\columnwidth]{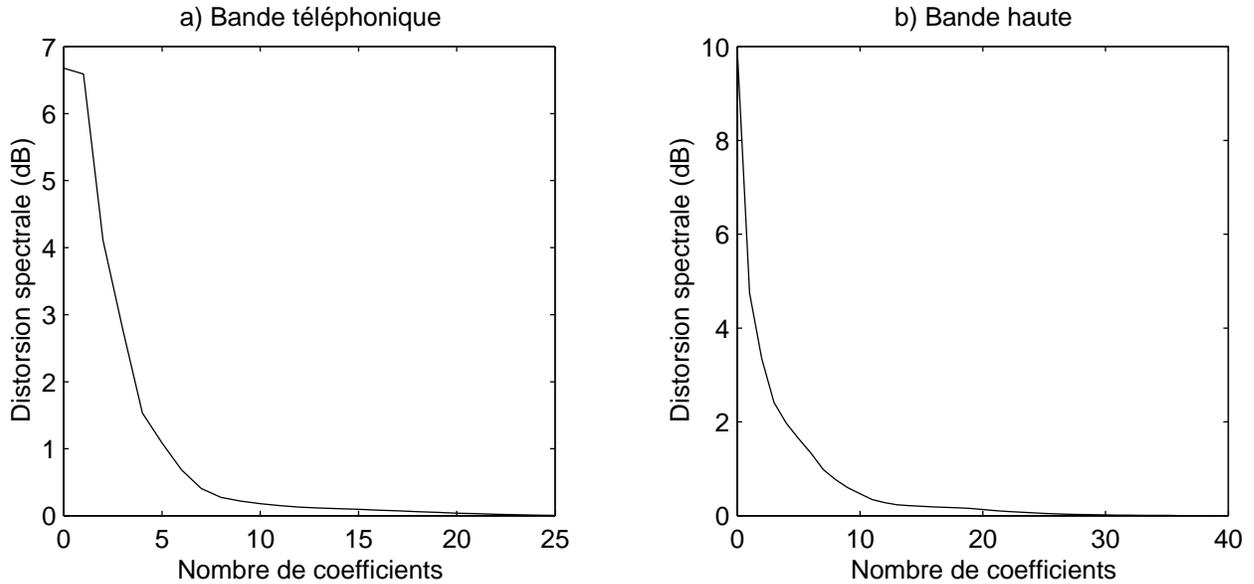}
\par\end{centering}

\caption[Distorsion spectrale et coefficients DCT]{Distorsion spectrale obtenue en fonction du nombre de coefficients de la DCT gardés (a) pour la bande téléphonique et (b) pour la bande haute.}\label{fig_dist_spectrale_DCT}
\end{figure}

Il reste ensuite à déterminer le nombre $N$ de coefficients de la
DCT qu'il sera nécessaire de garder. La figure \ref{fig_dist_spectrale_DCT}
montre la quantité d'information perdue (sous forme de distorsion
spectrale) en fonction du nombre de coefficients de la DCT que l'on
considère. Pour la bande téléphonique, on gardera 10 coefficients,
ce qui correspond à une distorsion spectrale de $0,2\:\mathrm{dB}$.
On voit clairement que 10 coefficients représentent presque parfaitement
l'enveloppe spectrale en bande téléphonique.

\begin{figure}[!t]
\begin{centering}
\includegraphics[width=1\columnwidth]{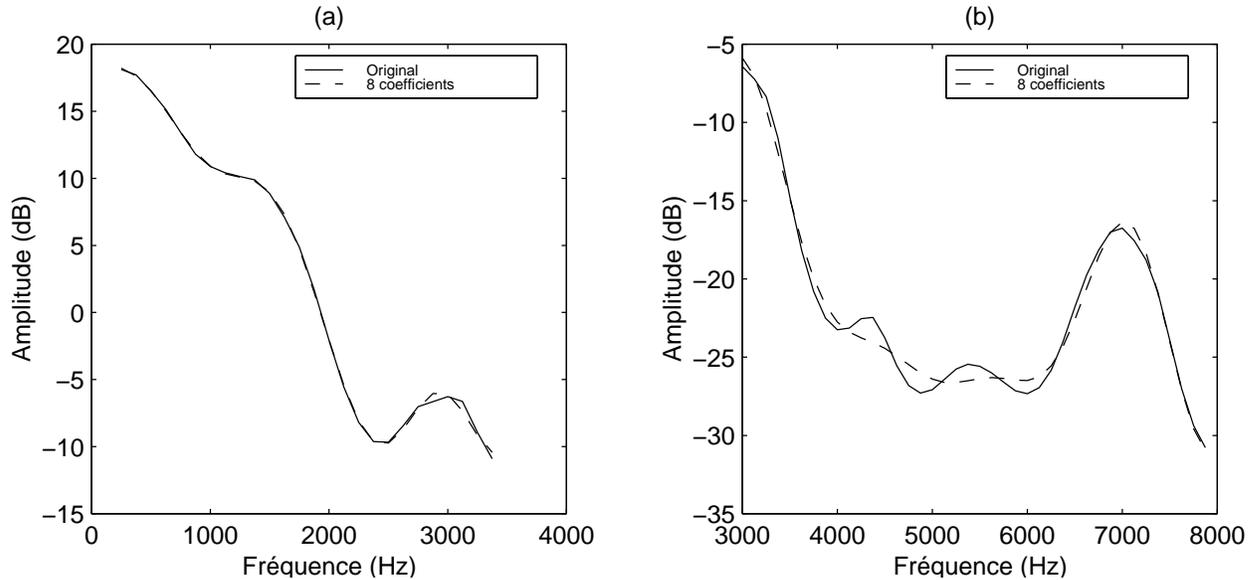}
\par\end{centering}

\caption[Exemple de spectre représenté par la DCT]{Enveloppe spectrale telle que représentée par (a) 10 coefficients dans le bande téléphonique (b) 8 coefficients dans la bande haute}\label{fig_exemple_DCT}
\end{figure}

Pour la bande haute, on gardera 8 coefficients, ce qui correspond
à une distorsion spectrale de $0,8\:\mathrm{dB}$. Bien que plus élevée
que pour la bande téléphonique, quelques tests perceptuels ont montré
que l'utilisation de plus de 8 coefficients de DCT n'améliore plus
la qualité sonore. La figure \ref{fig_exemple_DCT} montre, pour une
trame donnée la différence entre le spectre réel et le spectre tel
que représenté par la DCT avec un 10 coefficients pour la bande téléphonique
et 8 coefficients pour la bande haute.

Le nombre réduit de coefficients obtenu grâce à la DCT comporte plusieurs
avantages. D'abord, le temps de calcul est réduit, étant donné le
plus petit nombre de valeurs à traiter. Ensuite, certains systèmes
sont sensibles au nombre de dimensions impliquées dans le problème\footnote{Ce problème est connu sous le nom de \emph{curse of dimensionality}}.
C'est le cas des perceptrons multi-couches\cite{Haykin}, dont l'entraînement
se trouve facilité du nombre réduit de dimensions à traiter. 

\subsection{Extraction de paramètres vocaux\label{sec_feature_extraction}}

En plus de l'enveloppe spectrale dans la bande téléphonique, d'autres
caractéristiques de la voix peuvent être utiles afin de trouver l'enveloppe
spectrale aux hautes fréquences. Un de ces paramètres est le gain
de pitch. Celui-ci contient de l'information permettant de discriminer
facilement les voyelles des autres phonèmes. La période $T$ du pitch
elle-même peut aussi être utile pour son information sur le sexe du
locuteur, mais comporte certains inconvénients. En effet, la période
du pitch n'est définie que pour les trames voisées, ce qui a pour
conséquence que pour les trames non voisées ou peu voisées, la valeur
est mal définie. On verra plus loin que l'utilisation de la période
$T$ peut nuire à l'entraînement du système, selon le modèle utilisé.

En plus des paramètres reliés au pitch, on peut extraire de l'information
pertinente des variations temporelles du signal. On peut, par exemple
calculer la dérivée temporelle de l'énergie du signal et de la forme
du spectre. 

Au total, on choisit donc $17$ paramètres vocaux soit:
\begin{itemize}
\item $10$ coefficients cepstraux (DCT) dans la bande téléphonique (coefficients
$0$ à $9$)
\item Gain du pitch
\item Période du pitch
\item Dérivée temporelle de l'énergie de l'excitation (domaine logarithmique)
\item $4$ coefficients différentiels du cepstre (coefficients $0$ à $3$)
\end{itemize}
Le même vecteur de paramètres vocaux est utilisé pour l'entraînement
de tous les systèmes, sauf dans les cas ou un paramètre nuit visiblement
à l'entraînement (ce paramètre est alors ignoré).

\subsection{État de l'art}

L'extension hautes fréquences de l'enveloppe spectrale est, sans contredit,
la partie la plus complexe de ce projet. Elle nécessite en effet une
hétéro-association entre un vecteur de paramètres vocaux et le vecteur
de l'enveloppe spectrale haute. Les approches explorées jusqu'à maintenant
se divisent en trois grandes catégories, soit:
\begin{itemize}
\item Régressions\cite{Epps}
\item Dictionnaires associatifs (\emph{codebook mapping})\cite{Epps,Enbom}
\item Méthodes statistiques\cite{Ming,Jax}.
\end{itemize}
Pour ce qui suit, le vecteur de paramètres vocaux calculé sur la bande
téléphonique sera noté par $\mathbf{x}$ et le vecteur de la partie
haute de l'enveloppe spectrale sera noté par $\mathbf{y}$.

La méthode de la régression linéaire est la plus simple et consiste
à approximer le vecteur des hautes fréquences $\widehat{\mathbf{y}}$
par une combinaison linéaire des paramètres vocaux $\mathbf{x}$ en
bande téléphonique\footnote{Le vecteur $\mathbf{x}$ est augmenté du coefficient $1$ afin d'ajouter
un terme constant à la régression linéaire}, soit:

\begin{equation}
\widehat{\mathbf{y}}=\mathbf{Wx}\label{eq_reg_lin}
\end{equation}

où $\mathbf{W}$ est la matrice de régression. Soient $\mathbf{X}$
et $\mathbf{Y}$ des matrices contenant toutes les données d'entraînement
et dont les lignes sont respectivement l'ensemble des vecteurs $\mathbf{x}$
et \textbf{$\mathbf{y}$,} on désire trouver la matrice $\mathbf{W}$
qui minimise $\left\Vert \mathbf{Y}-\mathbf{WX}\right\Vert ^{2}$.
La solution peut être obtenue grâce à la méthode du pseudo-inverse,
soit: 
\begin{equation}
\mathbf{W}=\left(\mathbf{X}^{\mathrm{T}}\mathbf{X}\right)^{-1}\mathbf{XY}\label{eq_pseudo_inverse}
\end{equation}

L'article de Epps et Holmes\cite{Epps} montre que l'extension de
l'enveloppe spectrale par un dictionnaire associatif (\emph{codebook
mapping}) produit de meilleurs résultats que la régression linéaire,
surtout si l'on tient compte du gain de pitch. Un dictionnaire associatif
consiste en un quantificateur vectoriel sur des formes d'entrée (paramètres
vocaux) qui associe à chaque code un vecteur de sortie pour les hautes
fréquences \textbf{$\widehat{\mathbf{y}}$}. 

À la base d'un dictionnaire associatif se trouve un quantificateur
vectoriel sur l'ensemble de départ, soit les paramètres vocaux en
bande téléphonique. Le quantificateur vectoriel est alors utilisé
comme classificateur, séparant l'espace d'entrée en régions de Voronoï,
chacune ayant comme centroïde un des vecteurs représentant $\mathbf{x}_{i}$
du quantificateur. Le quantificateur vectoriel peut être entraîné
par l'algorithme LBG\cite{Gersho}.

\begin{figure}[!t]
\begin{centering}
\includegraphics[height=0.3\textwidth]{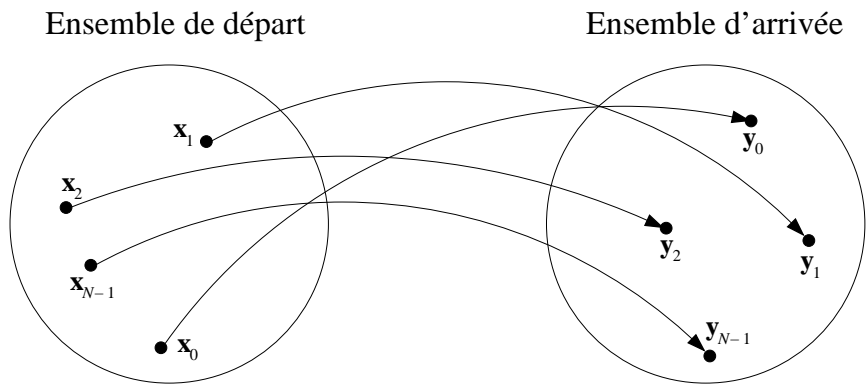}
\par\end{centering}

\caption{Dictionnaire associatif\label{fig_dict_associatif}}
\end{figure}

Chaque région de Voronoï, centrée sur un représentant $\mathbf{x}_{i}$
de l'ensemble d'entrée, est associée à un vecteur $\mathbf{y}_{i}$
de l'ensemble d'arrivée (fig. \ref{fig_dict_associatif}). Pour chaque
vecteur d'entrée $\mathbf{x}$, on trouve le centroïde $\mathbf{x}_{i}$
qui minimise la distance $\left(\mathbf{x}-\mathbf{x}_{i}\right)^{2}$
et on utilise $\mathbf{y}_{i}$ comme approximation du vecteur hautes
fréquences, soit:
\begin{equation}
\widehat{\mathbf{y}}=\mathbf{y}_{i},\qquad i=\min_{k}\:\left(\mathbf{x}-\mathbf{x}_{i}\right)^{2}\label{eq_codebook_yi}
\end{equation}

Les paramètres vocaux aux basses fréquences, utilisés comme \og clé \fg{}
du dictionnaire associatif, peuvent être des coefficients de l'enveloppe
spectrale elle-même\cite{Epps} ou des coefficients MFCC (\emph{Mel
frequency cepstral coefficients}) calculés sur l'enveloppe basses
fréquences ($200-3500\,\mathrm{Hz}$)\cite{Enbom}. De plus, il a
été montré que le fait d'utiliser l'information du pitch améliore
les résultats de l'extension\cite{Epps}. Dans le cas présent, on
utilise les mêmes vecteurs de paramètres vocaux que pour la régression,
à l'exception de la période du pitch $T$ et des coefficients cepstraux
différentiels qui nuisent à la performance du dictionnaire associatif.
Aussi, comme le quantificateur vectoriel utilise la mesure de distance
euclidienne et que le paramètre de gain de pitch $\beta$ est de faible
dynamique (généralement, $0\leq\beta\leq1$), on trouve que le fait
d'utiliser $4\beta$ comme paramètre produit de meilleurs résultats.

Des méthodes statistiques ont aussi été proposées pour résoudre le
problème de l'extension de l'enveloppe. Il s'agit soit de mixtures
de gaussiennes (GMM) \cite{Ming} soit de chaînes de Markov et de
GMM\cite{Jax}. 

La méthode statistique présentée par Ming, O'Shaughnessy et Mermelstein\cite{Ming}
consiste à quantifier les vecteurs d'enveloppe d'entrée et de sortie
et à calculer une fonction de probabilité $f\left(\mathbf{y}_{j}\left|\mathbf{x}_{i}\right.\right)$,
soit la probabilité d'observer la sortie (hautes fréquences) $\mathbf{y}_{j}$,
connaissant l'entrée (basses fréquences) $\mathbf{x}_{i}$. La fonction
$f\left(\mathbf{y}_{j}\left|\mathbf{x}_{i}\right.\right)$ est approximée
par une mixture de fonctions gaussiennes, soit:
\[
f=\sum_{i=0}^{N-1}a_{i}\exp\left[\frac{\left(\mathbf{x}-\overline{\mathbf{x}}_{i}\right)^{2}}{\sigma^{2}}\right].
\]

\subsection{Réseaux de neurones}

\begin{figure}[!t]
\begin{centering}
\includegraphics[height=0.2\textwidth]{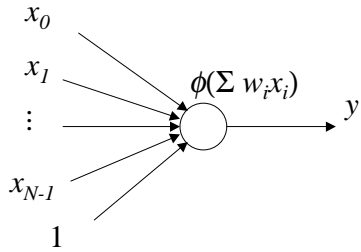}
\par\end{centering}

\caption[Unité d'un perceptron]{Neurone, unité de base d'un perceptron. La somme pondérée des entrées est calculée et une fonction d'activation non linéaire est appliquée au résultat pour calculer la sortie}\label{fig_def_neurone}
\end{figure}

Une autre méthode, entrant dans la catégorie des régressions non linéaires,
est utilisée ici pour l'extension de l'enveloppe spectrale. Il s'agit
des réseaux de neurones, ou plus précisément des perceptrons multi-couches.
Il est prouvé que les perceptrons utilisant plus d'une couche cachée
et respectant certaines contraintes entrent dans la catégorie des
approximateurs universels, c'est-à-dire qu'il peuvent approximer avec
une précision arbitraire n'importe quelle fonction\cite{Haykin}.
L'unité de base d'un perceptron multi-couches est appelée \og neurone \fg{}
ou parfois simplement \og unité \fg{}. Un neurone comporte plusieurs
entrées $x_{i}$ et une sortie $y$. La sortie $y$ est calculée en
appliquant une fonction non linéaire $\phi(x)$, nommée \og fonction
d'activation \fg{}, à la somme pondérée de ses entrées $x_{i}$ (fig.
\ref{fig_def_neurone}), soit:
\begin{equation}
y=\phi\left(\sum_{i=0}^{N-1}w_{i}x_{i}+b\right)\label{eq_def_neurone}
\end{equation}
où $\mathbf{w}$ est le vecteur de poids et la valeur $b$ est appelée
biais. Le biais peut-être considéré comme un poids dont l'entrée correspondante
est la constante 1. La fonction d'activation $\phi(x)$ la plus souvent
utilisée est la tangente hyperbolique:

\begin{equation}
\phi(x)=\tanh(x)=\frac{2}{1+\exp\left(-2x\right)}-1\label{eq_tangente_hyperbolique}
\end{equation}
Toutefois, comme on le verra plus loin, il est dans certains cas préférable
d'utiliser une fonction d'activation linéaire ($\phi(x)=x$). 

\begin{figure}[!t]
\begin{centering}
\includegraphics[height=0.3\textwidth]{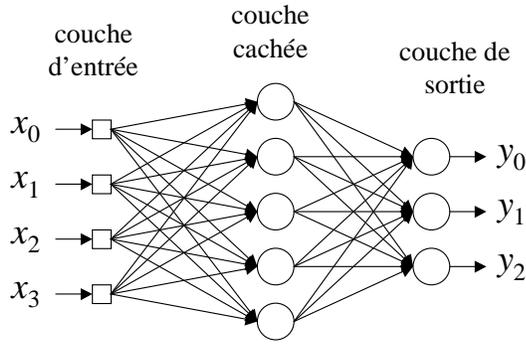}
\par\end{centering}

\caption[Perceptron multi-couches]{Exemple de perceptron à une couche cachée. Le perceptron  comprend 4 entrées, 5 unités sur la  couche cachée, et 3 sorties. Les différentes couches sont entièrement connectées.}\label{fig_perceptron}
\end{figure}

Un perceptron multi-couches est constitué d'un ensemble de neurones
arrangés sous forme de couches. Tous les neurones d'une couche sont
reliés à tous les neurones de la couche précédente, tel qu'illustré
à la figure \ref{fig_perceptron}. À chaque connexion et à chaque
biais est associé un poids $w_{i}$ et l'entraînement du perceptron
consiste à ajuster le vecteur de poids $\mathbf{w}$ de manière à
minimiser l'erreur entre la sortie du perceptron et la sortie désirée.
On minimise généralement la fonction d'erreur: 
\begin{equation}
E=\sum_{i=0}^{M-1}\sum_{j=0}^{N-1}\left(\widehat{y}_{ij}-y_{ij}\right)^{2}\label{eq_perceptron_erreur}
\end{equation}
où $M$ est le nombre d'exemples d'entraînement et $N$ est le nombre
de dimensions du vecteur de sortie\footnote{Dans le cas présent, $N=8$ car on garde $8$ coefficients de la DCT
pour la bande haute.}. 

L'entraînement d'un perceptron multi-couches devient donc un problème
d'optimisation du vecteur de poids $\mathbf{w}$. La méthode d'optimisation
la plus employée pour les perceptrons est appelé \og algorithme de
rétro-propagation \fg{} (\emph{back-propagation}) et est en fait
l'algorithme itératif de descente du gradient qui s'énonce:
\begin{equation}
w_{i}^{j+1}=w_{i}^{j}-\eta\frac{\partial E}{\partial w_{i}}\label{eq_back_propagation}
\end{equation}
où $w_{i}^{j}$ est le $i^{\grave{e}me}$ poids pour l'itération $j$
et $\eta$ est appelé \og pas d'apprentissage \fg{} (\emph{learning
rate}). Le pas d'apprentissage est un paramètre important de l'algorithme
de rétro-propagation. Plus $\eta$ est élevé, plus le réseau apprend
rapidement, mais une valeur trop grande rend l'algorithme d'entraînement
instable. 

En se référant à la figure \ref{fig_perceptron}, on voit que pour
la couche de sortie, pour un seul exemple d'entraînement donné:
\begin{equation}
\frac{\partial E}{\partial w_{ij}}=\frac{d\phi(z_{i})}{dz_{i}}h_{j}\left(\widehat{y}_{i}-y_{i}\right)\label{eq_gradient_sortie}
\end{equation}
où $w_{ij}$ est le poids pour l'entrée $j$ du neurone de sortie
$i$ et $z_{i}$ est la somme pondérée des entrées. Si la fonction
d'activation est la tangente hyperbolique, la dérivée s'exprime comme:

\begin{equation}
\frac{d\phi(z)}{dz}=\frac{1}{\cosh^{2}\:z}=1-\left[\phi(z)\right]^{2}\label{eq_derivee_tansig}
\end{equation}
ce qui facilite les calculs. Bien entendu, si la fonction d'activation
est linéaire, alors $d\phi(z)/dz=1$. Pour les neurones d'une couche
cachée, le principe demeure le même, alors que l'on applique la règle
de dérivation en chaîne pour trouver $\frac{\partial E}{\partial w}$.

Bien que très simple, l'algorithme de rétro-propagation a une convergence
très lente, ce qui rend peu intéressante son utilisation en pratique.
C'est pour cette raison qu'on utilise plutôt l'algorithme \emph{delta-bar-delta}
pour lequel un pas d'apprentissage $\eta_{i}$ est associé à chaque
poids $w_{i}$ du réseau et peut être adapté au cours de l'entraînement. 

Lors de l'entraînement, il est possible de constater que, comme suggéré
à la sous-section \ref{sec_feature_extraction}, l'utilisation de
la période du pitch $T$ nuit à la convergence de l'algorithme. Ce
paramètre est donc éliminé du vecteur de paramètres vocaux pour l'entraînement
des réseaux de neurones.

\section{Résultats}

L'ensemble d'entraînement comprend une heure et quarante minutes de
parole large bande sans les silences, alors que l'ensemble de validation
comprend quarante minutes de parole, toujours sans les silences.

\subsection{Régression linéaire}

Les résultats obtenus pour l'extension de l'enveloppe spectrale par
régression linéaire montrent une distorsion spectrale de $7,28\;\mathrm{dB}$
sur l'ensemble d'entraînement et de $7,24\;\mathrm{dB}$ sur l'ensemble
de validation. Cette méthode, bien que peu performante, est surtout
utilisée comme point de référence pour évaluer les résultats des autres
méthodes. Par exemple, un réseau de neurones qui n'atteindrait pas
au moins les performances de la régression linéaire sur l'ensemble
d'entraînement serait rejeté. Cela indiquerait en effet que l'optimisation
aurait convergé vers un minimum local bien loin du minimum global.

\subsection{Dictionnaires associatifs}

\begin{figure}[!t]
\begin{centering}
\includegraphics[height=0.3\textwidth]{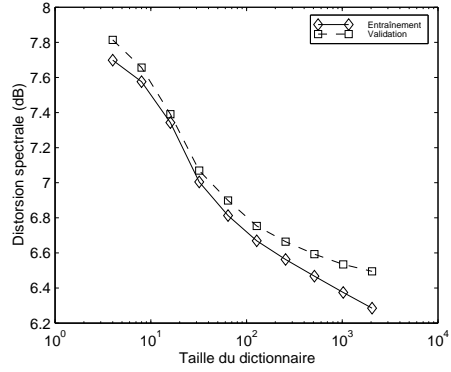}
\par\end{centering}

\caption[Performances des dictionnaire associatif]{Distorsion spectrale dans la bande haute sur l'ensemble d'entraînement et l'ensemble de validation en fonction de la dimension du dictionnaire associatif}\label{fig_perf_dict_accociatif}
\end{figure}

La figure \ref{fig_perf_dict_accociatif} montre les performances
d'extension de l'enveloppe pour différents dictionnaires associatifs
de taille allant de 4 à 2048 vecteurs. Avec plus de 2048 vecteurs,
le modèle devient trop complexe pour être implanté dans un système
fonctionnant en temps-réel. Il est tout de même clair que cette méthode
performe mieux qu'une simple régression linéaire.

\subsection{Perceptron multi-couches}

\begin{figure}[!t]
\begin{centering}
\includegraphics[height=0.3\textwidth]{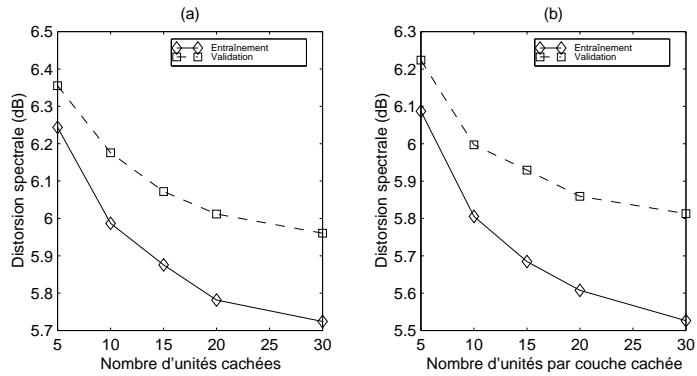}
\par\end{centering}

\caption[Hautes fréquences: performances des perceptrons multi-couches]{Performances des perceptrons multi-couches en fonction du nombre d'unités et de couches cachées}\label{fig_perf_perceptron}
\end{figure}

La figure \ref{fig_perf_perceptron} montre les performances des perceptrons
multi-couches. On remarque rapidement que même les perceptrons les
plus simples performent mieux que le dictionnaire associatif le plus
complexe évalué. On note toutefois qu'avec l'augmentation du nombre
d'unités cachées dans le perceptron, la différence entre l'erreur
sur l'ensemble d'entraînement et l'erreur sur l'ensemble de validation
s'accroît. Pour entraîner un perceptron avec plus de 30 unités cachées
par couche, il faudrait vraisemblablement augmenter la taille de l'ensemble
d'entraînement afin de réduire les effets de sur-généralisation.

\subsection{Analyse de complexité}

\begin{figure}[!t]
\begin{centering}
\includegraphics[height=0.3\textwidth]{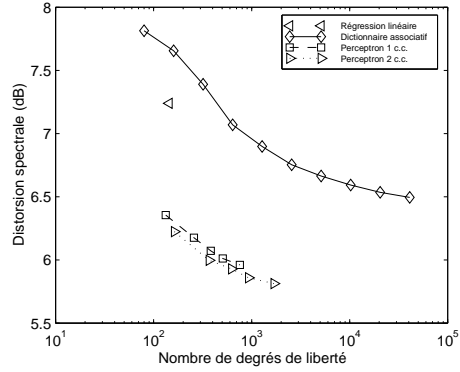}
\par\end{centering}

\caption[Performances vs. degrés de liberté]{Performances des différents modèles en fonction du nombre de degrés de liberté}\label{fig_complex_degres_liberte}
\end{figure}

\begin{figure}[!t]
\begin{centering}
\includegraphics[height=0.3\textwidth]{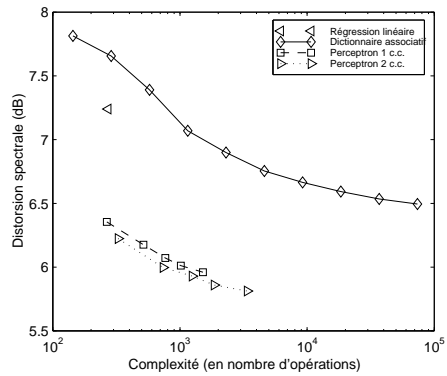}
\par\end{centering}

\caption[Performances vs. complexité algorithmique]{Performances des différents modèles en fonction de leur complexité algorithmique}\label{fig_complex_algo}
\end{figure}

Il est toujours possible de diminuer l'erreur de prédiction d'un modèle
sur son ensemble d'entraînement en complexifiant le modèle. Toutefois,
si on se réfère au rasoir d'Occam, les modèles simples sont souvent
les meilleurs\footnote{\og Les choses essentielles ne doivent pas être multipliées sans
nécessité \fg{}}. En général, un compromis est nécessaire entre la complexité d'un
modèle et la précision de la prédiction. En effet, un modèle trop
complexe comporte les inconvénients suivants: 
\begin{itemize}
\item mémoire nécessaire importante; 
\item temps de calcul important;
\item mauvaise généralisation.
\end{itemize}
Afin de contrôler en partie la capacité de généralisation d'un réseau
de neurones, on se réfère toujours à l'erreur de prédiction (distorsion
spectrale) sur l'ensemble de validation qui doit contenir des locuteurs
différents de l'ensemble d'entraînement. La figure \ref{fig_complex_degres_liberte}
montre l'erreur de prédiction obtenue en fonction de la complexité
du modèle, telle que donnée par le nombre de degrés de liberté du
modèle. La figure \ref{fig_complex_algo}, quant à elle, utilise le
nombre d'opérations par trame comme mesure de complexité.

On constate clairement qu'à complexité égale, les réseaux de neurones
performent beaucoup mieux que les dictionnaires associatifs. De plus,
les réseaux à deux couches cachées performent légèrement mieux que
ceux à une seule couche cachée, toujours à complexité égale.

\subsection{Reconstruction de la bande haute\label{sec_rec_acrtfacts}}

\begin{figure}[!t]
\begin{centering}
\includegraphics[width=1\columnwidth]{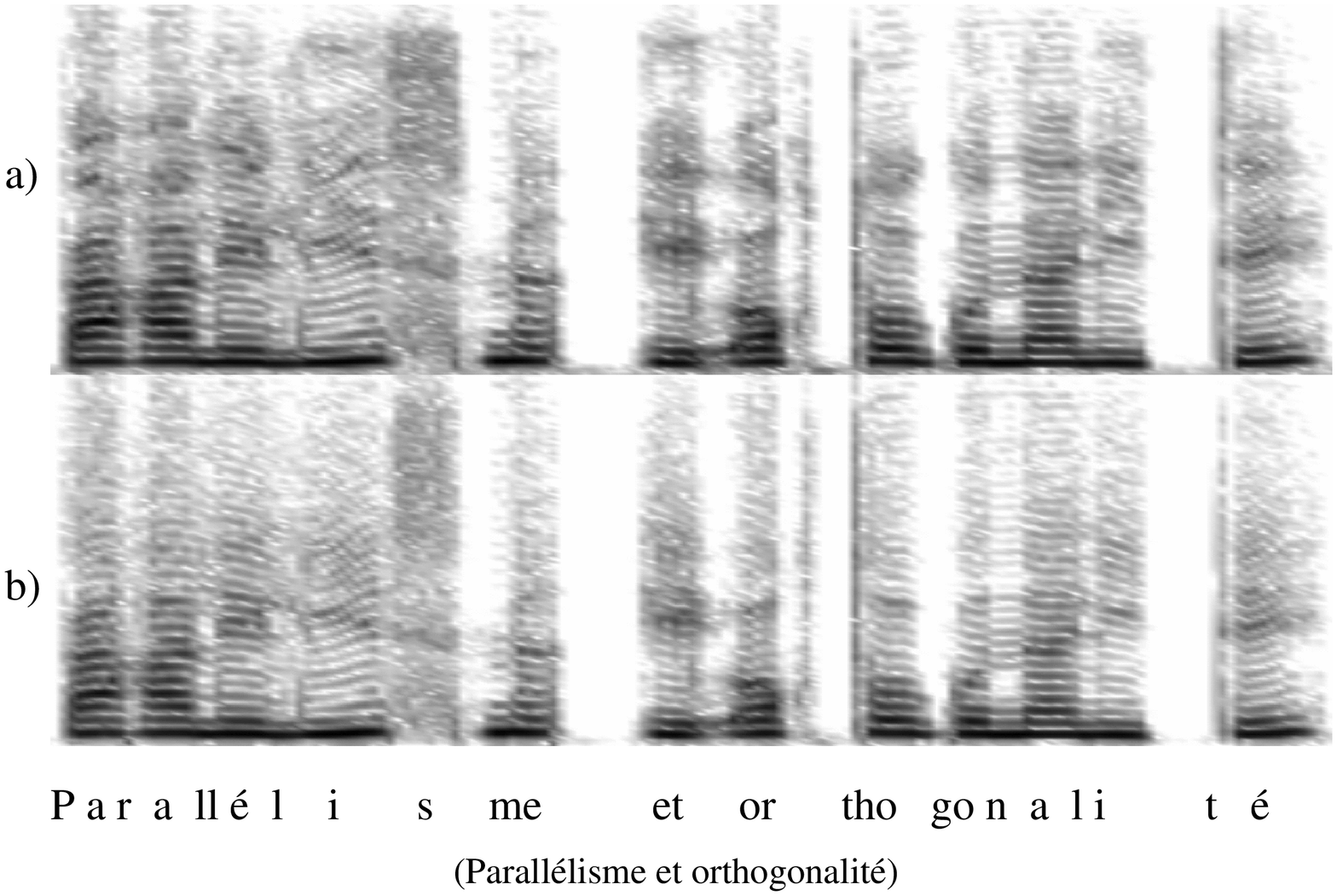}
\par\end{centering}

\caption[Spectrogramme: reconstruction de la bande haute]{Résultat de l'extension de la bande haute. Le spectrogramme du signal de parole reconstruit (a) est comparé à celui du signal dont seule l'excitation a été reconstruite (b) (envelope spectrale de l'original).}\label{fig_reconstruction_haute}
\end{figure}

Les spectrogrammes à la figure \ref{fig_reconstruction_haute} illustrent
bien l'effet de l'extension de l'enveloppe spectrale haute. Bien qu'en
général, l'enveloppe haute reconstruite suit bien l'enveloppe originale,
on note certains artefacts. Par exemple, on remarque que dans la reconstruction
le /s/ de \og parallélisme \fg{} manque un peu d'énergie. Ceci est
perçu lors de l'écoute comme un son à mi-chemin entre un /s/ et un
/$\int$/, ce qui est quelque peu désagréable.

De plus on note que certaines voyelles ont trop d'énergie en hautes
fréquences; c'est le cas du phonème /\emph{o}/ dans \og orthogonalité \fg{}
(le deuxième \og o \fg{}). Cet artefact est perçu comme une légère
distorsion dans le signal.

\section{Post-traitement}

Afin de réduire certains artefacts produits lors de l'extension, comme
ceux notés à la section \ref{sec_rec_acrtfacts}, on applique quelques
modifications au signal reconstruit. Ces modifications ont pour seul
but de rendre moins perceptibles ces artefacts.

\subsection{Lissage des variations dans l'enveloppe spectrale}

Afin de réduire certains artefacts désagréables aux hautes fréquences,
il est important de lisser les variations de l'enveloppe spectrale
dans les hautes fréquences. Ainsi, chaque coefficient du vecteur de
paramètres cepstraux de la bande haute est filtré par un FIR à trois
coefficients $H(z)=\frac{z+2+z^{-1}}{4}$. Ce filtre simple permet
d'éliminer la majorité des artefacts dus aux changements brusques
du spectre. Pour cette raison, il n'est pas nécessaire d'utiliser
des méthodes plus complexes comme les chaînes de Markov.

\subsection{Atténuation de la bande haute reconstruite}

Tous les systèmes explorés précédemment minimisent le critère d'erreur
quadratique (MMSE), ce qui fait que l'enveloppe spectrale estimée
est \og en moyenne \fg{} juste. Toutefois, dans la plupart des cas,
l'erreur perceptuelle causée par une sous-estimation de l'énergie
dans les hautes fréquences est beaucoup moins importante que celle
causée par une sur-estimation, cette dernière étant généralement perçue
comme du bruit ou de la distorsion. 

Ce sont justement les sur-estimation de l'enveloppe spectrale qui
affectent le plus la qualité du signal avec la bande haute reconstruite.
L'effet est encore plus marqué lorsque ces sur-estimations se trouvent
entre $3500\:\mathrm{Hz}$ et $4500\:\mathrm{Hz}$, c'est-à-dire dans
une bande où l'oreille est particulièrement sensible. 

Une manière d'améliorer la qualité perceptuelle du résultat de l'extension
des hautes fréquences consiste à atténuer \og artificiellement \fg{}
les hautes fréquences reconstruites. De plus, comme il a été constaté
que le fait de laisser un \og trou \fg{} dans la bande de $3500\:\mathrm{Hz}$
à $4500\:\mathrm{Hz}$ n'est que peu perceptible \cite{Jax}, cette
bande est coupée (fortement atténuée) de façon à réduire encore les
effets des erreurs sur l'enveloppe spectrale haute.

L'effet combiné de l'atténuation et du \og trou \fg{} dans la bande
élimine en bonne partie les artefacts causés lors de l'extension de
la bande haute.

\section{Codage de l'enveloppe\label{sec_codage_envelope}}

Pour certaines applications, soit celles où on code déjà le signal
en bande téléphonique, il peut être avantageux de \og dépenser \fg{}
quelques bits additionnels afin de corriger l'estimation de l'enveloppe
spectrale pour la bande haute. Ainsi, en quantifiant l'erreur de prédiction
de l'enveloppe spectrale, on peut réduire les artefacts dans la bande
haute\cite{Valin}. Par exemple, l'utilisation d'un quantificateur
vectoriel à $256$ entrées (soit $8\:\mathrm{bits}$) permet de réduire
la distorsion spectrale sur l'ensemble de validation de $5,81\:\mathrm{dB}$
à $1,83\:\mathrm{dB}$.

Si on n'utilise pas la prédiction et que l'on code directement la
bande haute sur $8\:\mathrm{bits}$, on obtient alors une distorsion
spectrale de $2,48\:\mathrm{dB}$. Il faut un quantificateur vectoriel
à $11\:\mathrm{bits}$ pour obtenir une distorsion spectrale équivalente
à celle que l'on obtient avec prédiction ($1,90\:\mathrm{dB}$ avec
$11\:\mathrm{bits}$). La prédiction par un réseau de neurones permet
donc d'économiser environ $3\:\mathrm{bits}$ d'information lors du
codage de l'enveloppe spectrale. Cette valeur peut être considérée
comme une approximation de l'information mutuelle entre la bande téléphonique
et la bande haute. Une autre évaluation de l'information mutuelle
est donnée dans \cite{Nilsson}.

\chapter{Extension des basses fréquences\label{sec_ext_BF}}

\section{Modèle utilisé}

Le problème de l'extension de la bande basse diffère de ce celui de
l'extension de la bande haute de trois principales façons: 
\begin{enumerate}
\item la largeur de bande à reconstruire pour la bande basse est beaucoup
plus petite ($\sim150\:\mathrm{Hz}$ contre $\sim3500\:\mathrm{Hz}$);
\item la sensibilité de l'oreille est beaucoup plus importante aux basses
fréquences;
\item le spectre aux basses fréquences est très harmonique.
\end{enumerate}
De plus, il a été observé que la majeure partie de l'information audible
aux basses fréquences est représentée seulement par les harmoniques
du pitch. On se préoccupe donc seulement de ces harmoniques. Pour
toutes ces raisons, le modèle sinusoïdal apparaît comme le plus approprié
pour l'extension des basses fréquences. Par modèle sinusoïdal, on
entend le fait de représenter un signal $x(t)$ (dans ce cas, les
basses fréquences reconstruites) par une somme de $N$ sinusoïdes,
soit:
\begin{equation}
x(n)=\sum_{i=0}^{N-1}A_{i}\sin\:\left(\omega_{i}n+\phi_{i}\right)\label{eq_modele_sinusoidal}
\end{equation}

Sachant que les harmoniques seront toujours des multiples de la fréquence
du pitch $\omega_{0}$ et que, même pour une voix grave ($\omega_{0}\sim70\:\mathrm{Hz}$),
on trouve au plus deux harmoniques dans la bande $50-200\:\mathrm{Hz}$,
on peut exprimer la bande basse reconstruite $x_{bf}(n)$ comme:
\begin{equation}
x_{bf}(n)=A_{0}\sin\:\left(\omega_{0}n+\phi_{0}\right)+A_{1}\sin\:\left(2\omega_{0}n+\phi_{1}\right)\label{eq_sinusoidal_params}
\end{equation}

Le système d'extension de la bande basse doit donc estimer 5 paramètres,
soit $A_{0}$, $\phi_{0}$, $A_{1}$, $\phi_{1}$, $\omega_{0}$.
On peut séparer le problème en deux parties:
\begin{enumerate}
\item synthèse de deux sinusoïdes d'amplitude unitaire aux bonnes fréquences
et phases;
\item amplification des sinusoïdes à un niveau se rapprochant le plus possible
de l'amplitude des sinusoïdes originales contenues dans le signal
large bande.
\end{enumerate}
Il faut noter que d'autres travaux sur l'extension de la bande basse
ont utilisé un modèle filtre-excitation, plutôt qu'un modèle sinusoïdal\cite{Milet}.
La raison principale du choix d'un modèle sinusoïdal ici est la constatation
que la partie stochastique du spectre sous $200\:\mathrm{Hz}$ est
très peu audible, ce qui rend le modèle sinusoïdal plus adapté (parce
que plus simple) que le modèle filtre-excitation.

\section{Synthèse des sinusoïdes\label{sec_BF_synthese}}

La fréquence du pitch $\omega_{0}$ utilisée pour les sinusoïdes peut
être facilement estimée à partir de la bande téléphonique. En effet,
soit $\omega_{0}=\frac{2\pi}{T}$, il s'agit d'effectuer une analyse
de pitch sur le signal en bande téléphonique. La méthode présentée
à la section \ref{sec_analyse_pitch} permet d'évaluer la période
$T$ du pitch. Malheureusement, l'application directe de l'équation
\ref{eq_pitch_T3} rend l'estimation du pitch sensible à un phénomène
appelé \emph{dédoublement du pitch}, où la période du pitch $T$ trouvée
est le double de la période réelle. Le dédoublement du pitch affecte
peu les codec de type CELP, qui compensent pour ces erreurs lors du
codage de l'excitation. Cependant, lors de l'extension des basses
fréquences, une erreur dans l'estimation du pitch se traduit directement
par une erreur dans la fréquence des harmoniques générés.

Le phénomène de dédoublement du pitch peut être partiellement enrayé
par l'utilisation d'une résolution temporelle plus élevée lors de
l'analyse\cite{Marques,Froon}. De plus, en ajoutant à l'algorithme
une partie logique favorisant les délais de pitch moins élevés, on
peut réduire encore les dédoublements du pitch.

\begin{figure}[!t]
\begin{centering}
\includegraphics[height=0.3\textwidth]{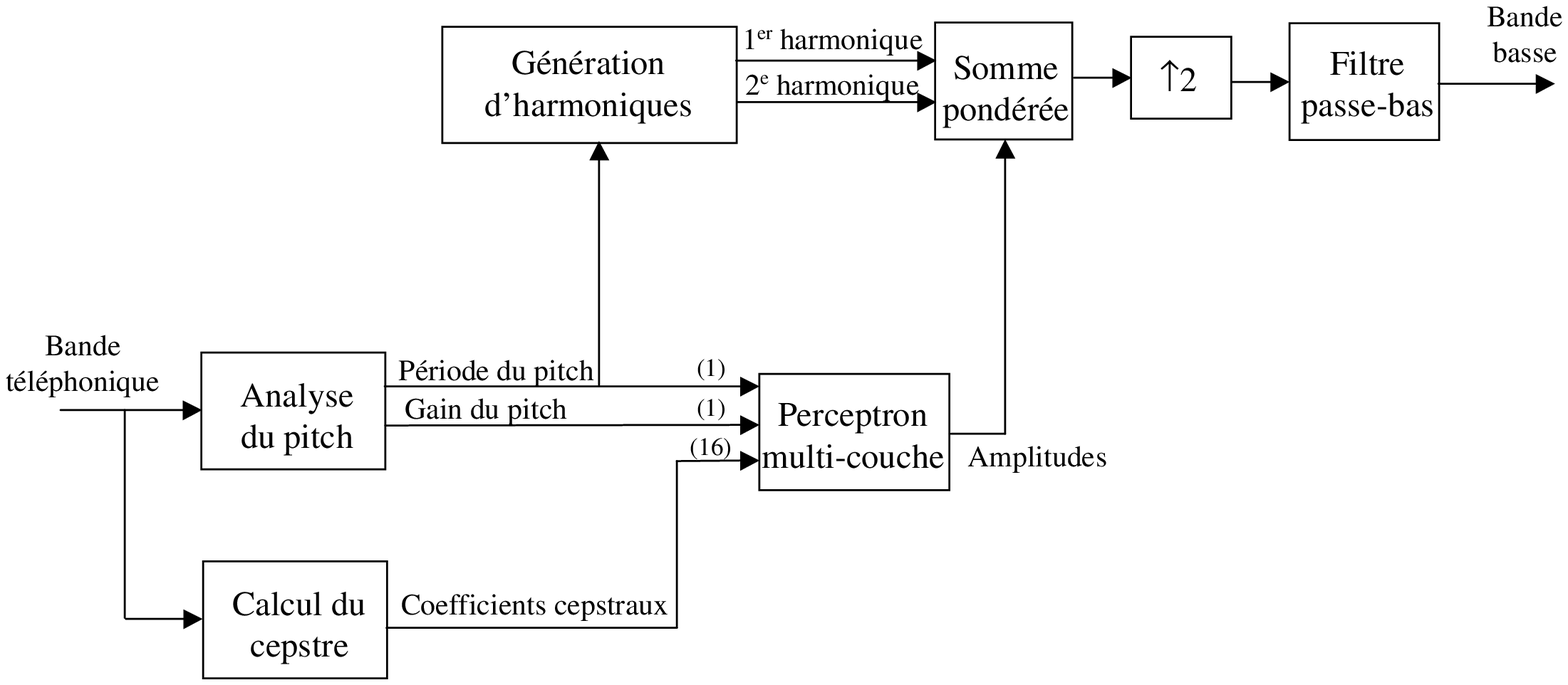}
\par\end{centering}

\caption[Système d'extension des basses fréquences]{Modèle sinusoïdal pour l'extension des basses fréquences}\label{fig_ext_bf1}
\end{figure}

Le problème des phases $\phi_{0}$ et $\phi_{1}$ est plus difficile
à résoudre. Toutefois, comme l'oreille ne peut percevoir les phases
de façon absolue, une première approximation consiste à imposer comme
seule contrainte la cohérence de $\phi_{0}$ et $\phi_{1}$ entre
les trames. Ainsi, on obtient deux générateurs sinusoïdaux modulés
en fréquence pour \og suivre \fg{} les deux premiers harmoniques
du pitch. Le système complet d'extension des basses fréquences est
donc celui montré à la figure \ref{fig_ext_bf1}.

\section{Estimation de l'amplitude des sinusoïdes\label{sec_BF_estimation_amplitude}}

Une fois les harmoniques générés, il est nécessaire d'estimer les
bonnes amplitudes. Tout comme pour l'extension des hautes fréquences
de l'enveloppe spectrale, il s'agit d'un problème d'hétéro-association
entre une forme d'entrée et le vecteur d'amplitude $\mathbf{a}=[A_{0}\,A_{1}]^{\mathrm{T}}$.
Afin de faciliter leur modélisation, le vecteur $\mathbf{a}$ est
normalisé par l'énergie de l'excitation et représenté dans le domaine
logarithmique, soit: 
\begin{equation}
\widetilde{\mathbf{a}}=\log\:\frac{\mathbf{a}}{\left\Vert r(n)\right\Vert }\label{eq_bf_amplitude_norm}
\end{equation}
où $r(n)$ est le signal d'excitation. 

Lors de l'entraînement il est nécessaire d'évaluer les amplitudes
normalisées $\widetilde{\mathbf{a}}$ à partir des données originales
large bande. Pour ce faire, le signal large bande est fenêtré Hanning
avec un recouvrement de 100 \%, de sorte à pouvoir utiliser le \og recouvrement
et addition \fg{} (\emph{overlap and add}). Ce signal est représenté
par des sinusoïdes fenêtrées de fréquences $\omega_{0}$ et $2\omega_{0}$.
Une trame fenêtrée $\mathbf{x}$ est approximée par:
\begin{equation}
\widetilde{\mathbf{x}}=\frac{1-\cos\:\frac{2\pi n}{N}}{2}\left(g_{0}+g_{1}\cos\:\omega_{0}n+h_{1}\sin\:\omega_{0}n+g_{2}\cos\:2\omega_{0}n+h_{2}\sin\:2\omega_{0}n\right)\label{eq_MSE_sine_fit}
\end{equation}
où les paramètres $g_{k}$ et $h_{k}$ sont obtenus par la méthode
des moindres carrés\footnote{On ne peut utiliser la transformée de Fourier, car le nombre de périodes
de pitch dans une trame n'est pas entier.}. Les amplitudes des sinusoïdes sont donc:
\begin{eqnarray}
A_{1} & = & \sqrt{g_{1}^{2}+h_{1}^{2}}\label{eq_amplitude_sqr1}\\
A_{2} & = & \sqrt{g_{2}^{2}+h_{2}^{2}}\label{eq_amplitude_sqr2}
\end{eqnarray}

Par souci de simplicité, l'extension des basses fréquences utilise
le même vecteur de paramètres vocaux que l'extension des hautes fréquences.
La méthode utilisée pour effectuer l'hétéro-association peut toutefois
être différente de celle utilisée pour la bande haute.

Des travaux comparant la méthode de la régression linéaire à celle
des dictionnaires associatifs\cite{Milet} pour l'extension de la
bande basse ont conclu que la première méthode produisait des résultats
significativement supérieurs à ceux de la seconde. Ces résultats concordent
avec certaines expériences préliminaires effectuées avec des dictionnaires
associatifs pour la bande basse. Pour ces raisons, seules deux méthodes
sont évaluées ici, soit la régression linéaire et les réseaux de neurones.

\section{Résultats quantitatifs}

Le critère d'erreur utilisé (minimisé) pour l'estimation des paramètres
basses-fréquences $A_{1}$ et $A_{2}$ est l'erreur quadratique moyenne
$E=\sqrt{\frac{A_{1}^{2}+A_{2}^{2}}{2}}$. Cette mesure est similaire
à la mesure de distorsion spectrale pour l'enveloppe haute fréquence.
Les données d'entraînement et de validation utilisées pour les basses
fréquences sont les mêmes que celles utilisées pour les hautes fréquences.

\subsection{Régression linéaire}

En utilisant la régression linéaire, on obtient une distorsion spectrale
de $3,65\;\mathrm{dB}$ sur l'ensemble d'entraînement et de $3,65\:\mathrm{dB}$
sur l'ensemble de validation. Ce résultat est comparable à ceux obtenus
par Milet \emph{et al.}\cite{Milet}, compte tenu des différences
pouvant exister entre différentes bases de données vocales.

\subsection{Réseaux de neurones}

\begin{figure}[!t]
\begin{centering}
\includegraphics[height=0.4\textwidth]{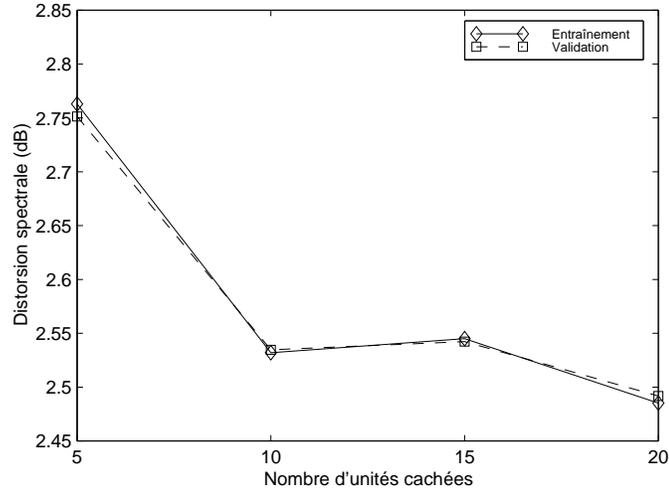}
\par\end{centering}

\caption[Basses fréquences: performances des perceptrons multi-couches]{Performances des perceptrons multi-couches en fonction du nombre d'unités cachées}\label{fig_results_bf_nn}
\end{figure}

Comme il n'a pas été possible de faire converger les réseaux de neurones
à deux couches cachées, seuls les résultats pour une seule couche
cachée sont présentés ici. La figure \ref{fig_results_bf_nn} montre
la distorsion spectrale obtenue pour différentes tailles de réseaux,
lors de l'entraînement et de la validation. Or, il semble que même
pour des perceptrons à une couche cachée, il soit difficile de bien
faire converger l'apprentissage. Ceci est mis en évidence par le fait
que les performances du perceptron à 10 unités cachées soit meilleures
que celles du perceptron à 15 unités cachées.

\section{Utilisation des basses fréquences résiduelles}

Bien que l'oreille soit insensible à la phase des signaux, la méthode
de synthèse des sinusoïdes présentée à la section \ref{sec_BF_synthese}
produit d'importants artefacts. Ceci est dû au fait que dans la bande
de transition entre la bande téléphonique et la bande basse reconstruite,
on additionne une partie d'un harmonique de pitch original au même
harmonique reconstruit avec une phase différente. Ceci crée des interférences
audibles entre les deux harmoniques.

Bien que l'on considère généralement que la bande téléphonique commence
à $200\:\mathrm{Hz}$, on remarque que dans la plupart des cas, la
bande $50-200\:\mathrm{Hz}$ n'est pas totalement atténuée. On retrouve
encore, à un niveau plus faible, l'information basse fréquence originale.
Il est donc intéressant de pouvoir récupérer cette information afin
d'améliorer la qualité du signal de sortie. En effet, même atténuée
de $20\:\mathrm{dB}$ ou plus, la bande basse contient toujours l'information
de phase des harmoniques du pitch. 

\begin{figure}[!t]
\begin{centering}
\includegraphics[height=0.3\textwidth]{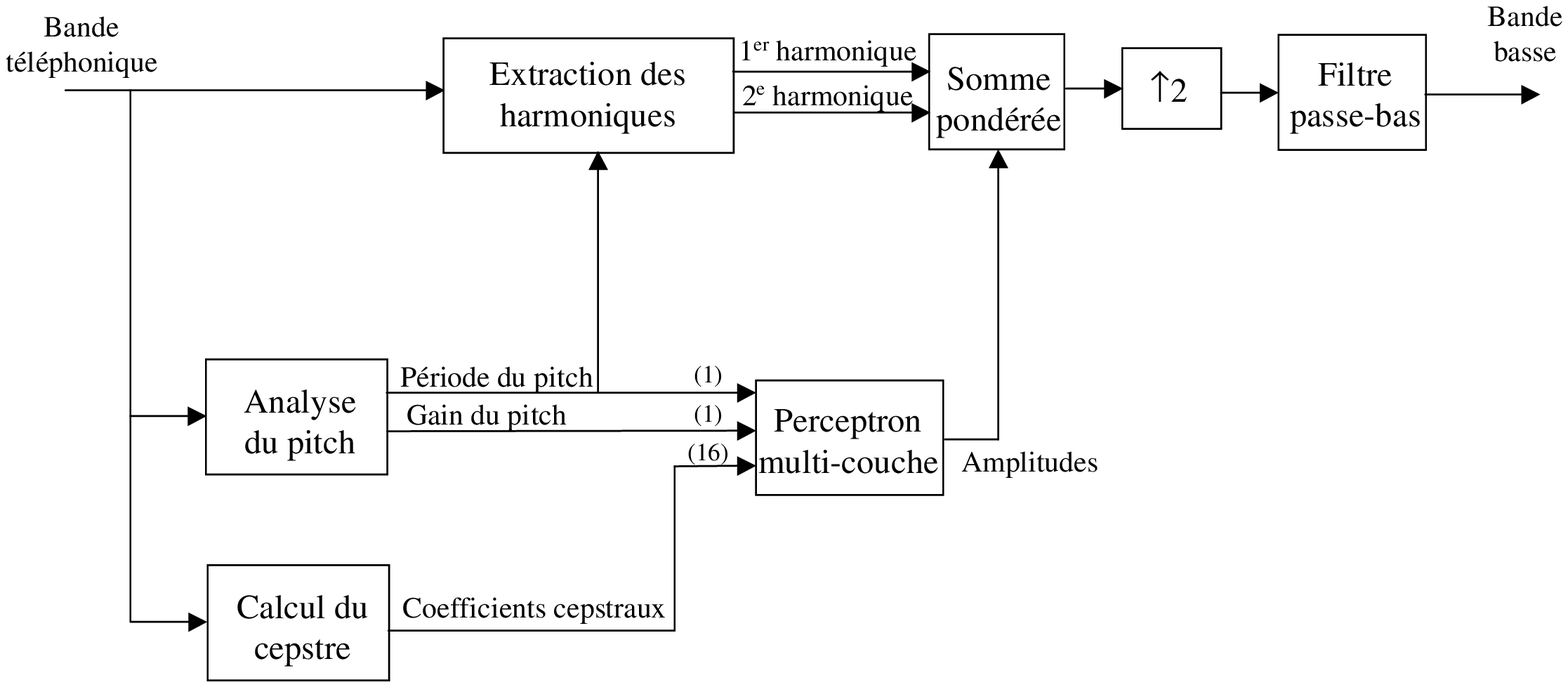}
\par\end{centering}

\caption[Système modifié d'extension des basses fréquences]{Modèle sinusoïdal pour l'extension des basses fréquences avec extraction des harmoniques résiduelles}\label{fig_ext_bf2}
\end{figure}

En calculant les paramètres $g_{k}$ et $h_{k}$ de l'équation \ref{eq_MSE_sine_fit}
à partir d'une trame en bande téléphonique, on obtient deux sinusoïdes
aux mêmes fréquences et phases que dans le signal original. Seules
les amplitudes sont corrigées en utilisant l'estimation obtenue par
la méthode choisie (section \ref{sec_BF_estimation_amplitude}). Cette
méthode réduit considérablement les artefacts dus aux erreurs de phase.
Le système d'extension des basses fréquences ainsi modifié est illustré
à la figure \ref{fig_ext_bf2}.

\chapter{Résultats\label{sec_resultats}}

\section{Résultats quantitatifs}

Le système d'extension de la bande comporte trois principales parties:
\begin{enumerate}
\item extension dans la bande basse (deux harmoniques);
\item extension de l'excitation dans la bande haute;
\item extension de l'enveloppe spectrale dans la bande haute.
\end{enumerate}
De ces trois parties, la deuxième, soit l'extension de l'excitation,
ne se prête pas à une évaluation quantitative. En effet, comme le
signal d'excitation comporte une grande partie aléatoire, on se contente
de reproduire un signal d'excitation \og plausible \fg{} qui soit
agréable pour l'oreille.

Les meilleurs résultats pour l'extension des basses fréquences sont
obtenus avec un réseau de neurones à une couche cachée comprenant
20 unités cachées. Dans ces conditions, l'erreur RMS est de seulement
$2,49\:\mathrm{dB}$ (figure \ref{fig_results_bf_nn}), ce qui est
de beaucoup inférieur à la valeur obtenue par régression linéaire,
soit $3,65\:\mathrm{dB}$.

L'extension de la bande haute est de loin la partie la plus difficile
à réaliser du système d'extension de la bande. En fait, la meilleure
méthode explorée produit une distorsion spectrale de $5,81\:\mathrm{dB}$
(figure \ref{fig_perf_perceptron}). Il s'agit d'un perceptron à deux
couches cachées avec 30 unités par couche cachée. Bien que la mesure
d'erreur utilisée pour les basses fréquences ne soit pas exactement
une mesure de distorsion spectrale, il est clair que l'erreur pour
les hautes fréquences est de beaucoup supérieure à celle pour les
basses fréquences ($5,81\:\mathrm{dB}$ pour les hautes fréquences,
comparé à $2,49\:\mathrm{dB}$ pour les basses fréquences).

\section{Résultats qualitatifs}

Les résultats quantitatifs sont utiles lors de la conception pour
comparer facilement différentes méthodes. Cependant, l'évaluation
finale de la qualité de l'extension doit se faire de façon qualitative
à partir de tests d'audition subjectifs.

\subsection{Méthode expérimentale}

La plupart des évaluations subjectives utilisées pour les codeurs
de parole utilisent des tests MOS. Malheureusement, ces tests sont
mal adaptés à une situation où l'on tente de comparer des échantillons
en bande téléphonique à des échantillons large bande. Il est donc
nécessaire de procéder à un autre type de test.

La méthode d'évaluation subjective retenue est la suivante. Pour chaque
phrase, l'auditeur doit comparer une paire de phrases traitée avec
le processus A à la même paire de phrases traitée avec le processus
B. Les traitements A et B sont choisis de façon aléatoire parmi trois
traitements possibles, soit:
\begin{enumerate}
\item aucun traitement, le signal large bande original est présenté;
\item le signal est filtré en bande téléphonique avec le filtre IRS modifié;
\item le signal large bande tel que produit par l'extension de la bande
à partir du signal en bande téléphonique.
\end{enumerate}
L'ordre dans lequel les fichiers sont présentés est choisi de façon
aléatoire. Pour chaque paire, l'auditeur doit répondre à la question:
\og Lequel des deux fichiers préférez-vous? \fg{}. Une question
aussi générale est choisie parce qu'il existe une différence importante
entre les trois différents types d'échantillons présentés. En effet,
il est nécessaire que l'écoute ne se fasse pas à partir de critères
précis comme la largeur de bande ou le niveau de distorsion, afin
de ne pas biaiser les résultats vers un traitement ou un autre.

En plus de choisir le \og meilleur \fg{} fichier, l'auditeur doit
aussi quantifier sa préférence sur une échelle de 1 à 5, où 1 signifie
\og très légère préférence \fg{} et 5 signifie \og très importante
préférence \fg{}.

Le même test est administré avec un casque d'écoute et avec des haut-parleurs.
Le test effectué avec casque d'écoute permet d'évaluer la qualité
sonore pour des applications de téléphonie. Le test avec haut-parleurs
vise à évaluer la qualité de l'extension pour des applications de
communications par Internet à partir d'un ordinateur personnel muni
de haut-parleurs.

\subsection{Résultats expérimentaux}

Les résultats des tests perceptuels pour casque d'écoute et haut-parleurs
sont présentés respectivement aux tableaux \ref{tab_tests_ec} et
\ref{tab_tests_hp}. Les résultats cumulatifs (casque d'écoute et
haut-parleurs ensemble) sont présentés au tableau \ref{tab_tests_both}.
Au total, 16 personnes ont participé aux tests avec casque d'écoute
et 13 personnes ont participé aux tests avec haut-parleurs\footnote{Ces deux ensembles ne sont pas disjoints}.

\begin{table}[!t]
\begin{centering}
\begin{tabular}{|c|c|c|c|}
\hline 
Casque d'écoute & Moyenne & Écart-type & Préférence (\%)\tabularnewline
\hline 
\hline 
Extension vs. bande téléphonique & 0,84 & 2.88 & 60\tabularnewline
\hline 
Large bande vs. extension & 1,92 & 1,91 & 79\tabularnewline
\hline 
Large bande vs. bande téléphonique & 2,84 & 1,64 & 88\tabularnewline
\hline 
\end{tabular}
\par\end{centering}

\caption{Résultats des tests subjectifs avec écouteurs\label{tab_tests_ec}}
\end{table}

\begin{table}[!t]
\begin{centering}
\begin{tabular}{|c|c|c|c|}
\hline 
Haut-parleurs & Moyenne & Écart-type & Préférence (\%)\tabularnewline
\hline 
\hline 
Extension vs. bande téléphonique & 0,51 & 2,92 & 59\tabularnewline
\hline 
Large bande vs. extension & 1,34 & 1,82 & 72\tabularnewline
\hline 
Large bande vs. bande téléphonique & 1,84 & 2,68 & 76\tabularnewline
\hline 
\end{tabular}
\par\end{centering}

\caption{Résultats des tests subjectifs avec haut-parleurs\label{tab_tests_hp}}
\end{table}

\begin{table}[!t]
\begin{centering}
\begin{tabular}{|c|c|c|c|}
\hline 
Résultats cumulés & Moyenne & Écart-type & Préférence (\%)\tabularnewline
\hline 
\hline 
Extension vs. bande téléphonique & 0,70 & 2,89 & 60\tabularnewline
\hline 
Large bande vs. extension & 1,68 & 1,88 & 76\tabularnewline
\hline 
Large bande vs. bande téléphonique & 2,38 & 2,23 & 83\tabularnewline
\hline 
\end{tabular}
\par\end{centering}

\caption{Résultats cumulatifs des tests subjectifs\label{tab_tests_both}}
\end{table}

\subsection{Analyse des résultats}

Pour l'analyse des résultats, comme les résultats avec casque d'écoute
sont très semblables aux résultats avec haut-parleurs (compte tenu
de marge d'erreur due au nombre de personnes ayant participé aux tests),
seuls les résultats combinés des deux expériences seront considérés
(tableau \ref{tab_tests_both}).

D'abord, si on ne considère que les préférences moyennes, on constate
que la préférence des signaux large bande, au détriment de la bande
téléphonique (2,38), est égale à la somme des deux autres résultats
($0,70+1,68=2,38$)\footnote{Le fait que le résultat soit parfaitement exact est toutefois l'effet
du hasard.}. On peut estimer que le processus d'extension de la bande retrouve
\emph{en moyenne} 30\% ($\frac{0,70}{2,38}=0,29$) de la qualité sonore
perdue lors du filtrage dans la bande téléphonique. Un résultat semblable
peut aussi être obtenu à partir des \og taux de préférence \fg{}
(dernière colonne du tableau \ref{tab_tests_both}). En effet, considérant
50\% comme taux de préférence neutre, on trouve $\frac{60\%-50\%}{83\%-50\%}=0,30$.

\begin{figure}[!t]
\begin{centering}
\includegraphics[height=0.4\textwidth]{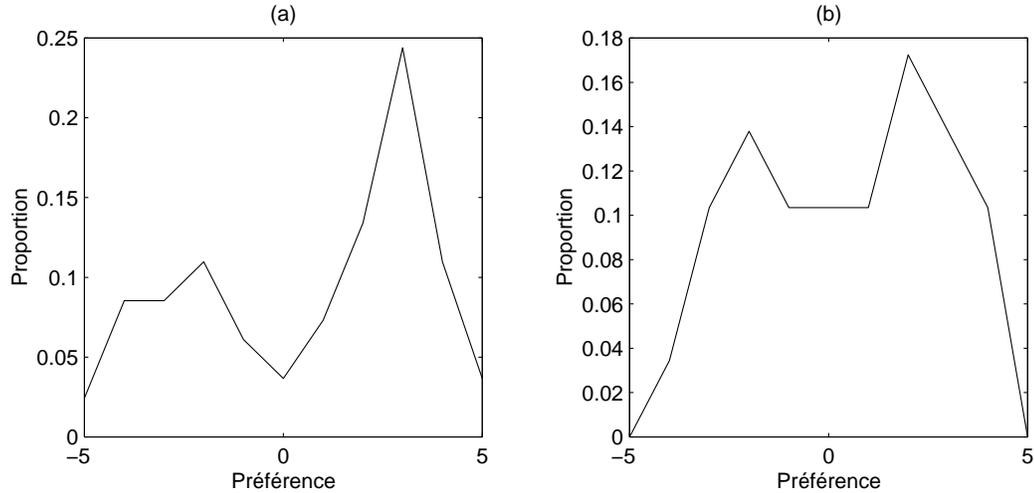}
\par\end{centering}

\caption[Distribution des préférences entre l'extension et la bande téléphonique]{Distribution des préférences entre l'extension et la bande téléphonique a) Distribution par fichier écouté b) Distribution par auditeur}\label{fig_results_distribution}
\end{figure}

Toutefois, en examinant les résultats, on constate que l'écart-type
pour la comparaison entre l'extension et la bande téléphonique est
très important, ce qui amène à examiner plus en détail la distribution
des résultats. Celle-ci est illustrée à la figure \ref{fig_results_distribution}.
Il est clair que la distribution des préférences ne suit pas la courbe
normale (gaussienne). On constate plutôt qu'il y a une importante
variabilité en fonction de l'auditeur. Ainsi, en observant les résultats
pour chaque auditeur, on constate que si le système d'extension avait
été en option sur un téléphone, environ la moitié aurait choisi de
l'utiliser\footnote{On considère ici les auditeurs pour lesquels la moyenne pour la comparaison
entre l'extension et la bande téléphonique est supérieure à zéro.}. On parle donc d'environ 50\% d'\emph{utilisateurs potentiels}. Le
tableau \ref{tab_results_utili} montre les résultats seulement pour
ces personnes. On note que ces utilisateurs potentiels:
\begin{itemize}
\item considèrent que l'extension se rapproche plus du large bande que de
la bande téléphonique;
\item apprécient plus la parole en large bande que la moyenne des participants
(large bande vs. bande téléphonique).
\end{itemize}
On peut donc supposer que pour environ la moitié des gens, la largeur
de bande est le principal facteur de qualité pour la parole, alors
que pour l'autre moitié des gens, l'important est de limiter le bruit
et la distorsion. En effet, d'après les commentaires de certains participants,
les effets de distorsion dans la bande haute constitue la principale
raison pour laquelle la bande téléphonique pouvait être préférée à
l'extension.

\begin{table}[!t]
\begin{centering}
\begin{tabular}{|c|c|c|c|}
\hline 
Utilisateurs potentiels & Moyenne & Écart-type & Préférence (\%)\tabularnewline
\hline 
\hline 
Extension vs. bande téléphonique & 2,52 & 1,68 & 91\tabularnewline
\hline 
Large bande vs. extension & 1,68 & 1,91 & 80\tabularnewline
\hline 
Large bande vs. bande téléphonique & 2,97 & 1,90 & 91\tabularnewline
\hline 
\end{tabular}
\par\end{centering}

\caption{Résultats pour les \emph{utilisateurs potentiels\label{tab_results_utili}}}
\end{table}

\chapter{Discussion et conclusion\label{sec_conclusion}}

Le système d'extension de la bande présenté permet de produire un
signal reconstruit en bande AM ($50\,\mathrm{Hz}$ à $7\,\mathrm{kHz}$)
à partir d'un signal d'entrée en bande téléphonique ($200\,\mathrm{Hz}$
à $3,5\,\mathrm{kHz}$). De plus, ce système peut fonctionner sur
un signal filtré \og IRS modifié \fg{}. 

L'extension est effectuée de façon indépendante pour les hautes fréquences
et les basses fréquences. La méthode retenue pour l'extension des
hautes fréquences (chapitre \ref{sec_ext_HF}) utilise le modèle filtre-excitation,
ce qui divise le problème en deux parties: l'extension de l'excitation
et de l'enveloppe spectrale. L'extension de l'excitation est réalisée
dans le domaine temporel par une une fonction non linéaire produisant
des harmoniques aux hautes fréquences. Cette méthode permet d'obtenir
un signal d'excitation presque impossible à distinguer de l'original.
C'est cette constatation qui rend la technique intéressante pour le
codage large bande à bas débit ou le codage \og encastré \fg{}.

L'extension de l'enveloppe spectrale haute est effectuée dans le domaine
cepstral. Plusieurs techniques ont été comparées, soit la régression
linéaire, les dictionnaires associatifs et les perceptrons multi-couches.
De ces trois techniques, les perceptrons produisent les meilleurs
résultats, bien que leur complexité soit relativement faible.

L'extension de la bande basse utilise le modèle sinusoïdal et ne reconstruit
que les deux premiers harmoniques du pitch. La fréquence de ces harmoniques
est estimée par une analyse du pitch, alors que la phase est estimée
à partir de l'information résiduelle dans la bande basse. Enfin, l'amplitude
des sinusoïdes est estimée par un perceptron multi-couches.

\section{Réalisation des objectifs initiaux}

Une liste des objectifs visés pour le système d'extension de la bande
est présentée à la section \ref{sec_objectifs_initiaux}. Voici le
degré d'atteinte de chacun de ces objectifs.

\subsection{Qualité sonore}

Les résultats des tests perceptuels montrent que les perceptions de
qualité de la parole après extension de la bande varie de façon importante
entre les différents participants. En effet, pour la moitié d'entre
eux, l'extension de la bande produit une nette amélioration de la
qualité alors que pour l'autre moitié, la qualité est réduite.

On pourrait donc envisager un service téléphonique ou un récepteur
pour lequel l'extension de la bande serait une option pouvant être
activée ou désactivée par l'utilisateur.

\subsection{Complexité}

La complexité totale du système proposé est estimée à environ 20 millions
d'opérations en virgule-flottante (simple précision) par seconde.
Ceci signifie qu'il est possible d'implanter l'algorithme sur un DSP.
De plus, il faut noter que les multiplications sont la plupart du
temps suivies d'une addition, ce qui améliore la performance sur la
plupart des DSP.

Sur la complexité de 20 millions d'opérations par seconde, environ
la moitié est due aux filtres. Tous les filtres utilisés sont des
FIR d'ordre élevé et de type non causal. Il serait donc possible de
réduire la complexité en optimisant mieux les filtres.

\subsection{Délai algorithmique}

Sauf lors de certaines analyses requérant l'utilisation de trames
avec recouvrement et dans certains filtres non causals, le système
d'extension décrit ici ne nécessite pas l'utilisation d'échantillons
futurs du signal. Le délai algorithmique est donc faible, soit de
l'ordre de deux trames, ce qui correspond à environ $32\:ms$. L'algorithme
d'extension peut donc être utilisé dans un système de transmission
de la parole en temps réel.

\section{Limitations et perspectives de recherche}

Le système d'extension de la bande présenté ici comporte certaines
faiblesses. La principale vient du fait qu'il est insensible au contexte.
Ainsi, seuls les paramètres de la trame courante sont utilisés pour
déterminer l'extension. Or, on sait qu'il est très difficile de faire
la reconnaissance de phonèmes de cette façon. C'est pourquoi les systèmes
de reconnaissance vocale actuels utilisent des mots entiers, voire
des phrases complètes pour effectuer la reconnaissance. 

Dans un contexte de communication en temps-réel, il n'est malheureusement
pas possible d'attendre la fin d'une phrase avant de faire l'extension.
Toutefois, il serait toujours possible d'utiliser l'information sur
le passé (système causal), et même ajouter un faible délai (moins
d'une seconde). 

Pour des applications sans contraintes de temps-réel tels des systèmes
de messagerie vocale, il serait possible d'effectuer la reconnaissance
de la parole sur chaque phrase et d'ensuite utiliser la segmentation
en phonèmes comme entrée supplémentaire au système d'extension. Ceci
permettrait entre autres de faire correctement l'extension des phonèmes
/s/ et /$\int$/, en utilisant le contexte. Par exemple, même si à
partir d'une trame isolée, on ne peut discriminer un /s/ d'un /$\int$/,
on sait que le mot \og statique \fg{} commence par un /s/ et non
par un /$\int$/.

Le post-traitement \og perceptuel \fg{} utilisé dans le système
d'extension pourrait aussi être amélioré. En effet, cette partie a
fait l'objet de relativement peu de recherche. Une des principales
améliorations potentielles consiste à appliquer un traitement différent
pour chaque classe phonétique (voyelles, fricatives, ...). Aussi,
il serait probablement possible de réduire la complexité du traitement,
notamment en optimisant les filtres utilisés et en utilisant un meilleur
algorithme de recherche du pitch.

Enfin, il serait intéressant d'évaluer plus en détail certaines applications
\og secondaires \fg{} de l'extension de la bande. Parmi ces applications,
on compte l'utilisation de l'extension de l'excitation afin de réduire
l'information transmise dans un codeur large bande. Aussi, certaines
méthodes de prédiction pourraient être employées pour réduire le débit,
comme présenté à la section \ref{sec_codage_envelope}.

\bibliographystyle{udsunsrt}
\bibliography{memoire}

\end{document}